\documentclass[onecolumn]{autart}

\usepackage{amsfonts}
\usepackage{amsmath}
\usepackage{amssymb}
\usepackage{array}
\usepackage{booktabs}
\usepackage{color}
\usepackage{comment}
\usepackage{caption}
\usepackage{geometry}
\usepackage{graphicx}
\usepackage{epstopdf}
\usepackage{float}
\usepackage{subcaption}
\usepackage{hyperref}

\numberwithin{equation}{section}

\nonstopmode

\def\ps@autart{%
 \def\@oddhead{}%
 \def\@evenhead{}%
 \def\@oddfoot{}%
 \def\@evenfoot{}%
}

\begin{document}

\title{A Quadratic Control Framework for Dynamic Systems}
\author{Igor Ladnik\thanks{\href{https://www.linkedin.com/in/igor-ladnik-ab93942}{LinkedIn}, \href{mailto:ladnik@hotmail.com}{email}}}
\date{}
\maketitle

\thispagestyle{empty}

\begin{abstract}
This article presents a unified approach to quadratic optimal control for both linear and nonlinear discrete-time systems, with a focus on trajectory tracking.
The control strategy is based on minimizing a quadratic cost function that penalizes deviations of system states and control inputs from their desired trajectories.

For linear systems, the classical Linear Quadratic Regulator (LQR) solution is derived using dynamic programming, resulting in recursive equations for feedback and feedforward terms. For nonlinear dynamics, the Iterative Linear Quadratic Regulator (iLQR) method is employed, which iteratively linearizes the system and solves a sequence of LQR problems to converge to an optimal policy.

To implement this approach, a software service was developed and tested on several canonical models, including: Rayleigh oscillator, inverted pendulum on a moving cart, two-link manipulator, and quadcopter.
The results confirm that iLQR enables efficient and accurate trajectory tracking in the presence of nonlinearities.

To further enhance performance, it can be seamlessly integrated with Model Predictive Control (MPC), enabling online adaptation and improved robustness to constraints and system uncertainties.
\end{abstract}

\newpage

\section{Introduction}

Quadratic optimal control remains one of the most fundamental and practically significant problems in control theory, with wide-ranging applications from robotics and aerospace systems to economics and biological motion modeling. Classical Linear Quadratic Regulator (LQR) theory provides an elegant closed-form solution for optimal control in linear systems with quadratic cost, offering both efficiency and robustness.

However, many real-world systems exhibit nonlinear dynamics, rendering direct application of LQR ineffective.
To address this, the iterative Linear Quadratic Regulator (iLQR) \cite{ilqr} method extends LQR to nonlinear domains through a sequence of linearizations and optimizations. This iterative approach preserves the computational efficiency of LQR while offering improved performance in nonlinear settings.

This article presents a unified treatment of both LQR and iLQR within a discrete-time trajectory tracking framework.
The cost function is expressed relative to desired trajectories for both states and controls.
The resulting method is applicable to a wide range of practical systems and is demonstrated through detailed simulations of
Rayleigh oscillator,
an inverted pendulum on a moving cart,
a two-link robotic manipulator and a quadcopter stabilization problem.

In addition, the iLQR method can be efficiently combined with \textit{Model Predictive Control} (MPC). MPC is an advanced control technique that solves a finite-horizon optimization problem at each time step using the current system state as the initial condition. By integrating iLQR into the MPC framework, one can rapidly compute optimal control sequences for nonlinear systems while continuously updating the control policy based on real-time feedback. This hybrid iLQR–MPC scheme is particularly advantageous for systems with constraints, disturbances, or time-varying dynamics, where both predictive adaptation and computational tractability are critical.

\section{Problem Statement}

\begin{figure}[H]
    \centering
\begin{minipage}[b]{0.45\linewidth}
   \centering
     \includegraphics[width=\linewidth]{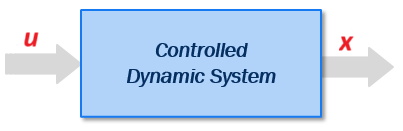}
   \end{minipage}
     \caption{Open-loop control system.}
     \label{fig:Open-loop}
\end{figure}

We begin by formulating the optimal control problem for a nonlinear dynamical system governed by the differential equation:
\begin{equation}
\dot{x}(t) = f(t, x(t), u(t)), \qquad t \in [0, T],
\label{eq:2-1}
\end{equation}
where $x(t) \in \mathbb{R}^n$ is the system state, and $u(t) \in \mathbb{R}^m$ is the control input (Figure~\ref{fig:Open-loop}).

\noindent
Our objective is to minimize the following quadratic cost function:
\begin{equation}
\begin{aligned}
J(u) = \int_0^T \Big[ & (\hat{x}(t) - x(t))^\top  Q(t) (\hat{x}(t) - x(t)) + (\hat{u}(t) - u(t))^\top  R(t) (\hat{u}(t) - u(t)) \Big] dt,
\end{aligned}
\label{eq:2-2}
\end{equation}
where $\hat{x}(t)$ and $\hat{u}(t)$ denote the desired state and control trajectories. The matrices $Q(t)$ and $R(t)$ are symmetric and time-dependent, with $Q(t)$ positive semi-definite and $R(t)$ positive definite.

This formulation corresponds to a standard \textit{trajectory tracking problem}, commonly encountered in control theory. The goal is to determine an optimal control input $u(t)$ such that the system follows the desired trajectories as closely as possible while minimizing the cost.

In the following sections, we first examine the linear case, deriving the optimal solution using dynamic programming in discrete time. We then extend this approach to nonlinear systems using an iterative linearization technique, corresponding to the iLQR method \cite{ilqr}.

\newpage

\section{Linear Case}

First, we consider a \textit{linear dynamic system} governed by the well-known differential equation:
\begin{equation*}
    \dot{x}(t) = \bar{A}(t)\,x(t) + \bar{B}(t)\,u(t)
\end{equation*}
where $x(t)$ is the state vector, $u(t)$ is the control input, and $\bar{A}(t)$ and $\bar{B}(t)$ are the system and input matrices, respectively.

In discrete form, this equation can be approximated as:
\begin{equation}
    x_{k+1} = A_k\,x_k + B_k\,u_k
\label{eq:3-1}
\end{equation}
with the discrete-time matrices $A_k$ and $B_k$ defined by:
\begin{equation*}
\begin{split}
    A_k &\approx I + \bar{A}(k\Delta t)\,\Delta t \\
    B_k &\approx \bar{B}(k\Delta t)\,\Delta t
\end{split}
\end{equation*}
where $I$ is the identity matrix and $\Delta t$ is the discretization time step.

\noindent
Suppose that $\hat{x}_0=x_0$.
Then the discretization of the cost function \eqref{eq:2-2} can be presented with the following expression
\begin{equation}
    J_N(u) = \sum_{k=0}^{N-1}[(\hat{x}_{k+1} - x_{k+1})^\top Q_k(\hat{x}_{k+1} - x_{k+1}) + (\hat{u}_k - u_k)^\top R_k(\hat{u}_k - u_k)]
\label{eq:3-2}
\end{equation}

\noindent
Let's denote minimum of \(J_N(u)\) as
\begin{equation*}
    m_N(x_0) = \min_{(u_0,u_1,...,u_{N-1})} J_N(u)
\end{equation*}
and the minimum of the cost function on the last $N-(k+1)$ steps of the process is $m_{N-(k+1)}(x_{k+1})$.

\noindent
Consider the $k$-th step of the process.
Assuming that the control over the last $N - (k + 1)$ steps is optimal, and applying Bellman's principle of optimality, the following relation is obtained:
\begin{equation}
\begin{split}
    m_{N-k}(x_k) = \min_{u_k}\big[(\hat{x}_{k+1} - x_{k+1})^\top Q_k(\hat{x}_{k+1} - x_{k+1}) + (\hat{u}_k - u_k)^\top R_k(\hat{u}_k - u_k) + m_{N-(k+1)}(x_{k+1})\big]
\end{split}
\label{eq:3-3}
\end{equation}
where $k = 0, 1,2, ..., N-1$.

\noindent
At $k = N$, we have $m_0(x_N) = 0$.

\noindent
Let’s assume that $m_{N-k}(x_k)$ can be presented as
\begin{equation}
    m_{N-k}(x_k) = x_k^\top P_{N-k}x_k - 2v_{N-k}^\top x_k + a_{N-k}
\label{eq:3-4}
\end{equation}
where
$P$ is symmetric positively semi-defined matrix, \(v\) is vector and \(a\) is a value independent on \(x\).

\noindent
Replacing in \eqref{eq:3-3} \(m_{N-(k+1)}(x_{k+1})\) with expression \eqref{eq:3-4} for $(k+1)$-th step, getting
\begin{equation}
\begin{split}
    m_{N-k}(x_k) = \min_{u_k}\big[&(\hat{x}_{k+1} - x_{k+1})^\top Q_k(\hat{x}_{k+1} - x_{k+1}) + (\hat{u}_k - u_k)^\top R_k(\hat{u}_k - u_k) + \\
    &x_{k+1}^\top P_{N-(k+1)}x_{k+1} - 2v_{N-(k+1)}^\top x_{k+1} + a_{N-(k+1)}\big]
\end{split}
\label{eq:3-5}
\end{equation}

\noindent
The control sequence $u_k$ is determined by substituting the expression for $x_{k+1}$ from equation \eqref{eq:3-1} into equation \eqref{eq:3-5}, and setting to zero the partial derivative with respect to $u_k$ of the expression in square brackets in \eqref{eq:3-5}.
\begin{equation}
    u_k = c_{N-k} - F_{N-k}x_k
\label{eq:3-6}
\end{equation}
where
\begin{equation}
\begin{split}
    &W_{N-k} = B_k^\top  (Q_k + P_{N-(k+1)}) B_k + R_k \\[1ex]
    &c_{N-k} = W_{N-k}^{-1}\left[B_k^\top (Q_k\hat{x}_{k+1} + v_{N-(k+1)}) + R_k\hat{u}_k\right] \\[1ex]
    &F_{N-k} = W_{N-k}^{-1}\left[B_k^\top (Q_k + P_{N-(k+1)})A_k\right]
\end{split}
\label{eq:3-7}
\end{equation}
\noindent
If matrix  $W$ is near-singular, then a product of the regularization parameter (as in Levenberg–Marquardt regularization) and the identity matrix may be added to it.

Substituting \eqref{eq:3-6}, \eqref{eq:3-7} and \eqref{eq:3-1} into \eqref{eq:3-5},
an expression for $m_{N-k}(x_k)$ is derived in the form corresponding to \eqref{eq:3-4}, where
\begin{equation}
\begin{split}
&Z_k = A_k - B_kF_{N-k} \\[1ex]
&P_{N-k} = Z_k^\top (Q_k + P_{N-(k+1)})Z_k + F^\top _{N-k}R_kF_{N-k} \\[1ex]
&v_{N-k} = Z_k^\top (Q_k \hat{x}_{k+1} + v_{N-(k+1)} - (Q_k + P_{N-(k+1)}) B_k c_{N-k}) + F_{N-k}^\top R_k (c_{N-k} - \hat{u}_k)
\end{split}
\label{eq:3-8}
\end{equation}

\begin{figure}[H]
    \centering
\begin{minipage}[b]{0.45\linewidth}
   \centering
     \includegraphics[width=\linewidth]{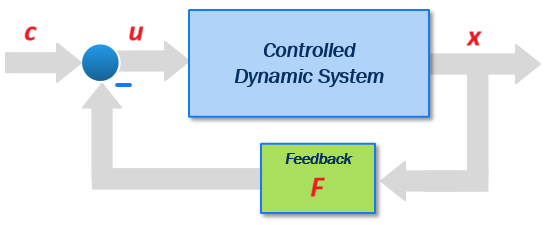}
   \end{minipage}
     \caption{Closed-loop control system.}
     \label{fig:Closed-loop}
\end{figure}

\newpage
\noindent
Expressions \eqref{eq:3-6}--\eqref{eq:3-8} define the optimal control of the open-loop system $u_k$, as illustrated in Figure~\ref{fig:Closed-loop}, where $c_{N-k}$ represents the control vector of the closed-loop system, and $F_{N-k}$ is the feedback gain matrix applied to all coordinates of the state vector.
Both of these parameters are generally time-dependent.

Theoretically, different combinations of $c_{N-k}$ and $F_{N-k}$ can yield the same values of $u_k$ in equation \eqref{eq:3-6}. However, such variations can significantly affect the system's response to unmodeled disturbances.

Clearly, not all coordinates of the state vector are directly measurable. The unmeasured components are estimated using various observers, such as the Kalman filter.

\begin{figure}[H]
    \centering
\begin{minipage}[b]{0.7\linewidth}
   \centering
     \includegraphics[width=\linewidth]{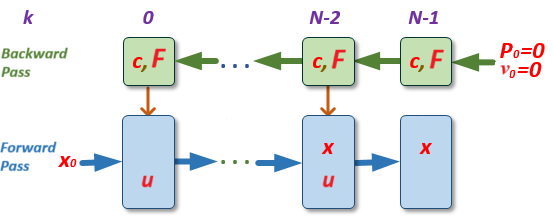}
   \end{minipage}
     \caption{Computation schema.}
     \label{fig:Computation_schema}
\end{figure}

The computation process proceeds as follows: First, during the backward pass, the parameters $c_{N-k}$ and $F_{N-k}$ are computed using equations \eqref{eq:3-7} and \eqref{eq:3-8}, starting from the final time step.
Next, in the forward pass, the control vector $u_k$ is computed using equation \eqref{eq:3-6}, and the corresponding state vector is then updated via the system dynamics described in equation \eqref{eq:2-1}.
The overall procedure is illustrated in Figure~\ref{fig:Computation_schema}.

\newpage

\section{Nonlinear Case}

Now we extend the above result to the nonlinear systems. The main idea is using the above recurrent mechanism, to obtain control policy with iterations for linearized control object. For that matter we linearize discrete representation of \eqref{eq:2-1}
\begin{equation}
	x_{k+1} = f(k,\ x_k,\ u_k)
\label{eq:4-1}
\end{equation}

\noindent
For \(i\)-th iteration (trajectory) the controlled object is presented in the following form:
\begin{equation}
    x_{k+1}^{i+1} = A_k^ix_k^{i+1} + B_k^iu_k^{i+1}
\label{eq:4-2}
\end{equation}

\noindent
where transfer matrices at the $k$-th point of $i$-th trajectory are given with the following Jacobians
\begin{equation}
\begin{aligned}
A_k^i &= \nabla_x f(k, x_k^i, u_k^i) = \left. \frac{\partial f}{\partial x} \right|_{x_k^i, u_k^i}, \\
B_k^i &= \nabla_u f(k, x_k^i, u_k^i) = \left. \frac{\partial f}{\partial u} \right|_{x_k^i, u_k^i}.
\label{eq:4-3}
\end{aligned}
\end{equation}

\noindent
State and control vectors on \((i+1)\)-th trajectory may be presented as:
\begin{equation}
\begin{split}
    &x_k^{i+1} = x_k^i + \delta{x}_k^i \\
    &u_k^{i+1} = u_k^i + \delta{u}_k^i
\end{split}
\label{eq:4-4}
\end{equation}
\noindent
Cost function \eqref{eq:3-2} may be formulated for increments $\delta{x}$ and $\delta{u}$.
In this case, in the cost function \eqref{eq:3-2}, vectors of desirable values $\hat{x}$ and $\hat{u}$ should be replaced with new vectors $\delta{\hat{x}}$ and $\delta{\hat{u}}$ respectively:
\begin{equation}
\begin{split}
    &\delta{\hat{x}_{k+1}^{i+1}} = \hat{x}_{k+1} - x_{k+1}^i \\
    &\delta{\hat{u}_k^{i+1}} = \hat{u}_k - u_k^i
\end{split}
\label{eq:4-5}
\end{equation}
\noindent
This allows us to reduce the nonlinear problem to a sequence of linear iterations.

\section{Computational Algorithm for Nonlinear Case}

The algorithm consists of the following steps:

\noindent
\\ \emph{1}. Assume initial (for iteration $0$) vectors $x^0_k$ and $u^0_k$ (normally all zero values except for initial conditions $x^0_{k=0}$, but the choice is up to a user).

\noindent
\\ \emph{2}. Calculate linearized $A^i_k$ and $B^i_k$ for all $k$ at the $i$-th trajectory according to \eqref{eq:4-3}.
This can be performed either analytically, by deriving Jacobians, or numerically using finite differences or automatic differentiation.

\noindent
\\ \emph{3}. Calculate new vectors $\delta{\hat{x}_{k+1}^i}$ and $\delta{\hat{u}_k^i}$ of desirable values for the $i$-th iteration according to \eqref{eq:4-5}.

\noindent
\\ \emph{4}. Calculate $\delta{x}^i_k$ and $\delta{u}^i_k$ with inverse and direct runs as for the linear case, replacing in \eqref{eq:3-6}--\eqref{eq:3-8}, and \eqref{eq:3-1}
actual and desirable state and control vectors with their respective increments:
\begin{equation*}
    x_k \Rightarrow \delta{x}^i_k,\quad
    u_k \Rightarrow \delta{u}^i_k,\quad
    \hat{x}_k \Rightarrow \delta{\hat{x}}^i_k,\quad
    \hat{u}_k \Rightarrow \delta{\hat{u}}^i_k.\quad
\end{equation*}
Regularization may be be applied in this step to ensure numerical stability.

\noindent
\\ \emph{5}. Calculate $x^{i+1}_k$ and $u^{i+1}_k$ for $(i+1)$-th iteration with \eqref{eq:4-4}.

\noindent
\\ \emph{6}. Check the condition to stop iterations.
This condition may be based, e.g., on the difference between values of the cost function compared to the previous iteration, or simply contain a fixed number of iterations.
If the condition is not satisfied, then steps \emph{2 - 6} should be repeated until the stop condition will be satisfied.
After the condition was satisfied, policy $u_k$ and states $x_k$ is considered as the final solution.

\noindent
\\ It is important to note that even in the nonlinear case, when using a single iLQR iteration with constant
desired state $\hat{x}$, control vector $\hat{u}$, and weight matrices $Q$ and $R$, both the input vector $c$ and the feedback matrix $F$ in equation \eqref{eq:3-6} remain constant.
This observation enables the system designer to simplify, in some cases, the closed-loop control of nonlinear systems by reducing it to time-invariant input and feedback terms.

Several numerical examples illustrating different scenarios are presented below.

\newpage

\section{Numeric Examples}

The quadratic optimization algorithm was implemented as a software service and tested on several canonical models of dynamic systems.

\subsection{Rayleigh Oscillator}

This dynamic system is described as following \cite{rayleigh1, rayleigh2}.

\begin{table}[H]
\centering
\caption{State and Control Variables}
\label{tab:rayleigh_vars}
\begin{tabular}{llcll}
\toprule
& \# & \textbf{Variable} & \textbf{Description} \\
\midrule
\textbf{State Vector}
& 1 & $x_0$ & Position \\
& 2 & $x_1$ & Velocity \\
\midrule
\textbf{Control Input}
& 1 & $u$   & Control input \\
\bottomrule
\end{tabular}
\end{table}

\begin{table}[H]
\centering
\caption{System Parameters and Equations}
\label{tab:rayleigh_symbols}
\begin{tabular}{ll}
\toprule
\textbf{Symbol} & \textbf{Description} \\
\midrule
$a$ & System nonlinearity parameter ($a = 1.4$) \\
$b$ & Control gain ($b = 4$) \\
\midrule
$\dot{x}_0$ & $x_1$ \\
$\dot{x}_1$ & $-x_0 + a (1 - \frac{x_1^2}{10}) x_1 + b u$ \\
\bottomrule
\end{tabular}
\end{table}

\noindent
Initial state of the system is presented in the Table~\ref{tab:rayleigh_xInit}.

\begin{table}[H]
\centering
\small
\caption{Initial State Vector}
\label{tab:rayleigh_xInit}
\begin{tabular}{>{\bfseries}lcc}
\toprule
Parameter & $x$ (m) & $v_x$ (m/s) \\
\midrule
Value     & -5       & -5 \\
\bottomrule
\end{tabular}
\end{table}

\noindent
The objective of the control in this case is to drive the system to the zero state with minimal control effort.
Constant desired values of state and control and their weight factors are presented in the Table~\ref{tab:rayleigh_desired}.

\begin{table}[H]
\centering
\caption{Desired Values and Weights}
\label{tab:rayleigh_desired}
\begin{tabular}{c *{2}{c} @{\hspace{6pt}}|@{\hspace{6pt}} c}
\toprule
\textbf{Variable} & $x_0$ & $x_1$ & $u$ \\
\midrule
\textbf{Desired Value} & 0 & 0 & 0 \\
\textbf{Weight}        & 1 & 0 & 1 \\
\bottomrule
\end{tabular}
\end{table}

\noindent
Two cases of control design are shown below.
In the left-hand side of Figure~\ref{fig:rayleigh_comparison} a single iteration variant of control is presented.
In this case the control may be implemented with constant negative feedbacks (Table~\ref{tab:rayleigh_const}) without input of close-loop system (since the steady state is zero).
The three iterations case is presented in the right-hand side of Figure~\ref{fig:rayleigh_comparison}.
Although the dynamics of the system state coordinates are similar in both cases, the initial control signal in the open-loop system with constant feedback gains is significantly higher—approximately 9 vs.\ 6~m/s\textsuperscript{2}.

\newpage

\begin{table}[H]
\captionsetup{justification=raggedright, singlelinecheck=false}
\caption{The Single Iteration Case}
\label{tab:rayleigh_const}
\begin{tabular}{ll}
\toprule
\textbf{Symbol} & \textbf{Value} \\
\midrule
$c$ (closed-loop input) & $0$ \\
\midrule
$F$ (feedback matrix) &
$\begin{bmatrix}
0.7591 & 1.064
\end{bmatrix}$ \\
\bottomrule
\end{tabular}
\end{table}

\begin{figure}[htbp]
    \centering

    \begin{minipage}[b]{0.48\textwidth}
        \centering
        \includegraphics[width=\textwidth]{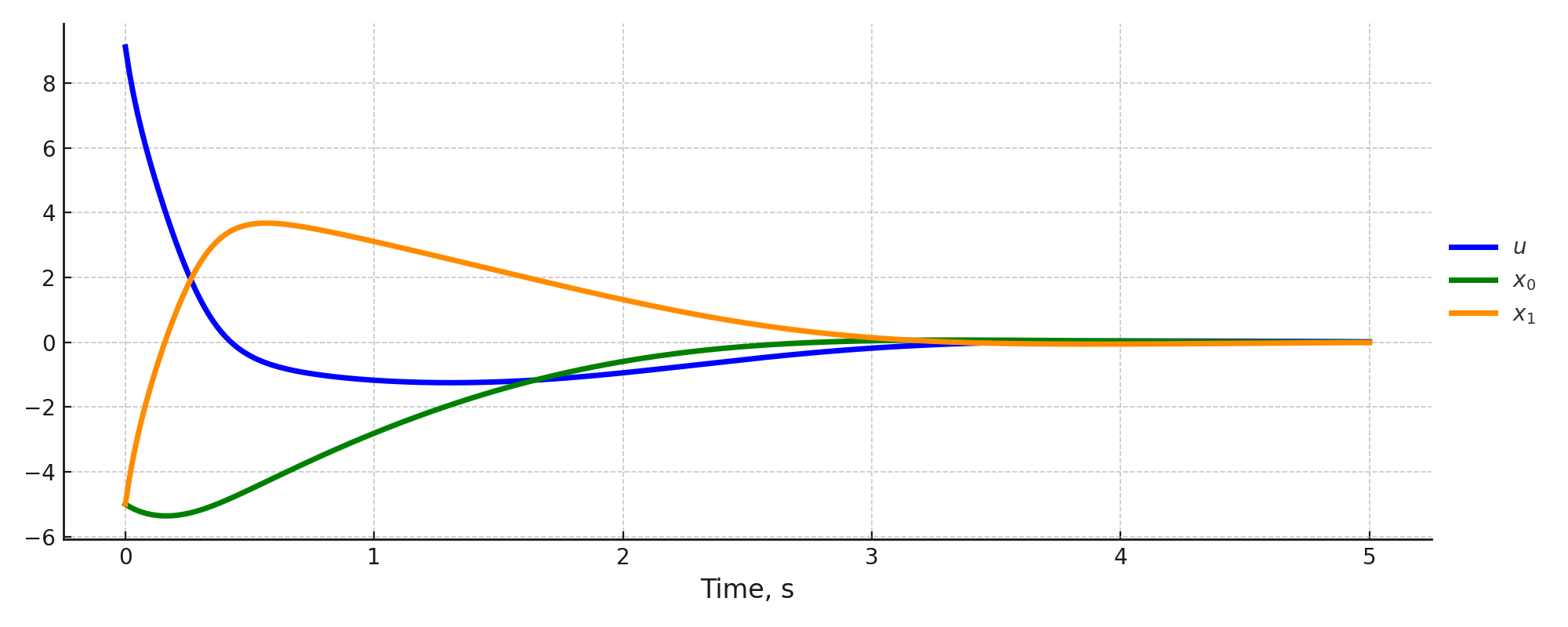}
        \caption*{1 iteration}
    \end{minipage}
    \hfill
    \begin{minipage}[b]{0.48\textwidth}
        \centering
        \includegraphics[width=\textwidth]{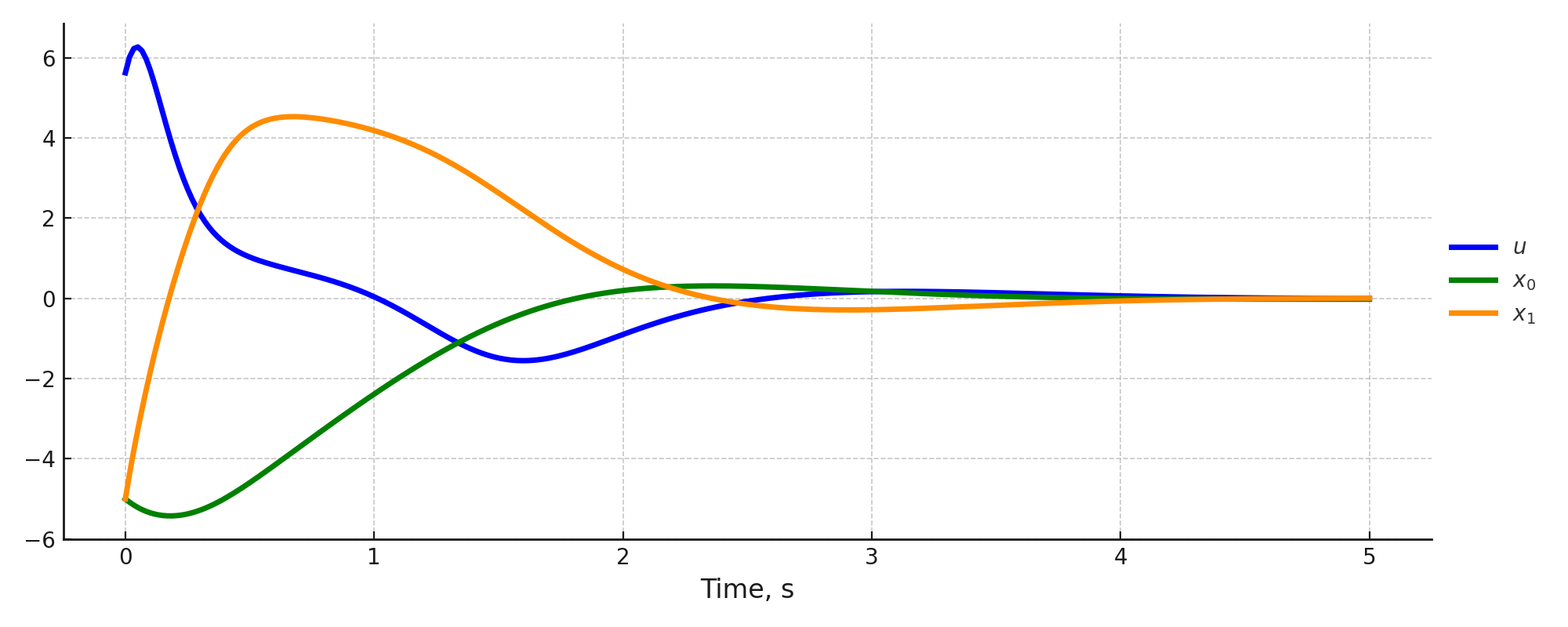}
        \caption*{3 iterations}
    \end{minipage}

    \caption{Comparison of Rayleigh system dynamics under different scenarios.}
    \label{fig:rayleigh_comparison}
\end{figure}

\subsection{Inverted Pendulum on a Moving Cart}

\begin{figure}[H]
    \centering
    \begin{minipage}[b]{0.45\linewidth}
    \centering
    \includegraphics[width=\linewidth]{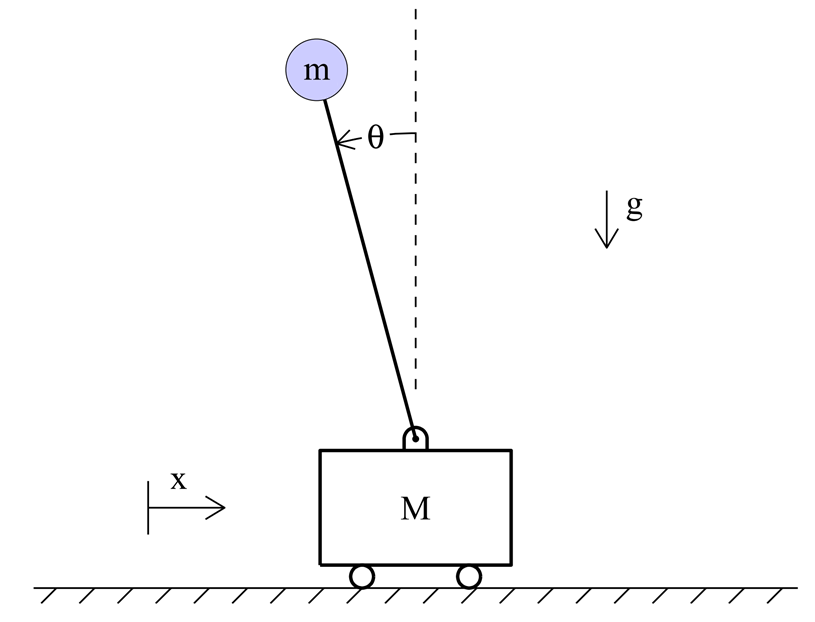}
    \end{minipage}
    \caption{Schema of Inverted Pendulum on a Moving Cart \cite{pendulum1}.}
    \label{fig:Pendulum_schema}
\end{figure}

\vspace{1em}
The inverted pendulum on a moving cart (Figure~\ref{fig:Pendulum_schema}) is a classic nonlinear control system consisting of a pendulum hinged on a horizontally moving cart \cite{pendulum1, pendulum2}.
The goal is to apply a horizontal force to the cart in order to stabilize the pendulum in its upright (inverted) position, despite its inherent instability.
This system often serves as a benchmark problem for validating control algorithms in robotics, automation, and feedback system design.

\begin{table}[H]
\centering
\caption{State and Control Variables}
\label{tab:pendulum_vars}
\begin{tabular}{llcll}
\toprule
& \# & \textbf{Variable} & \textbf{Units} & \textbf{Description} \\
\midrule
\textbf{State Vector} & 1 & $x$ & m & Cart position \\
& 2 & $\dot{x}$ & m/s & Cart velocity \\
& 3 & $\theta$ & \textdegree & Pendulum angle from vertical \\
& 4 & $\dot{\theta}$ & \textdegree/s & Angular velocity of pendulum \\
\midrule
\textbf{Control Input} & 1 & $u$ & N & Horizontal force on the cart \\
\bottomrule
\end{tabular}
\end{table}

\begin{table}[H]
\centering
\caption{System Dynamics and Parameters}
\label{tab:pendulum_symbol}
\begin{tabular}{ll}
\toprule
\textbf{Symbol} & \textbf{Description} \\
\midrule
$g$ & Acceleration due to gravity (9.81 m/s\textsuperscript{2}) \\
$m$ & Mass of the pendulum (1 kg) \\
$M$ & Mass of the cart (1 kg) \\
$l$ & Length to pendulum center of mass (1 m) \\
\midrule
$\dot{x}$ & Cart velocity \\
$\dot{\theta}$ & Angular velocity of pendulum \\
$\ddot{x}$ & $\frac{-m l \dot{\theta}^2 \sin\theta + m g \sin\theta \cos\theta + u}{M + m(1 -\cos^2\theta)}$ \\
$\ddot{\theta}$ & $\frac{- m l \dot{\theta}^2 \sin\theta \cos\theta + (M + m) g \sin\theta + u \cos\theta}{l ({M + m(1 -\cos^2\theta)}}$ \\
\bottomrule
\end{tabular}
\end{table}

\noindent
The control objective is to shift the cart by 10 meters while stabilizing the pendulum in the upright position.
The corresponding constant desired values for the state and control vectors, along with their weights, are provided in Table~\ref{tab:pendulum_desired}.

\begin{table}[H]
\centering
\caption{Desired Values and Weights for Inverted Pendulum}
\label{tab:pendulum_desired}
\begin{tabular}{c *{4}{c} @{\hspace{6pt}}|@{\hspace{6pt}} *{2}{c}}
\toprule
\textbf{Variable} & $x$ & $\dot{x}$ & $\theta$ & $\dot{\theta}$ & $u$ \\
\midrule
\textbf{Desired Value} & 10 & 0 & 0 & 0 & 0 \\
\textbf{Weight}        & 100 & 1 & 1 & 1 & 10 \\
\bottomrule
\end{tabular}
\end{table}

\begin{figure}[htbp]
    \centering

    \begin{minipage}[b]{0.48\textwidth}
        \centering
        \includegraphics[width=\textwidth]{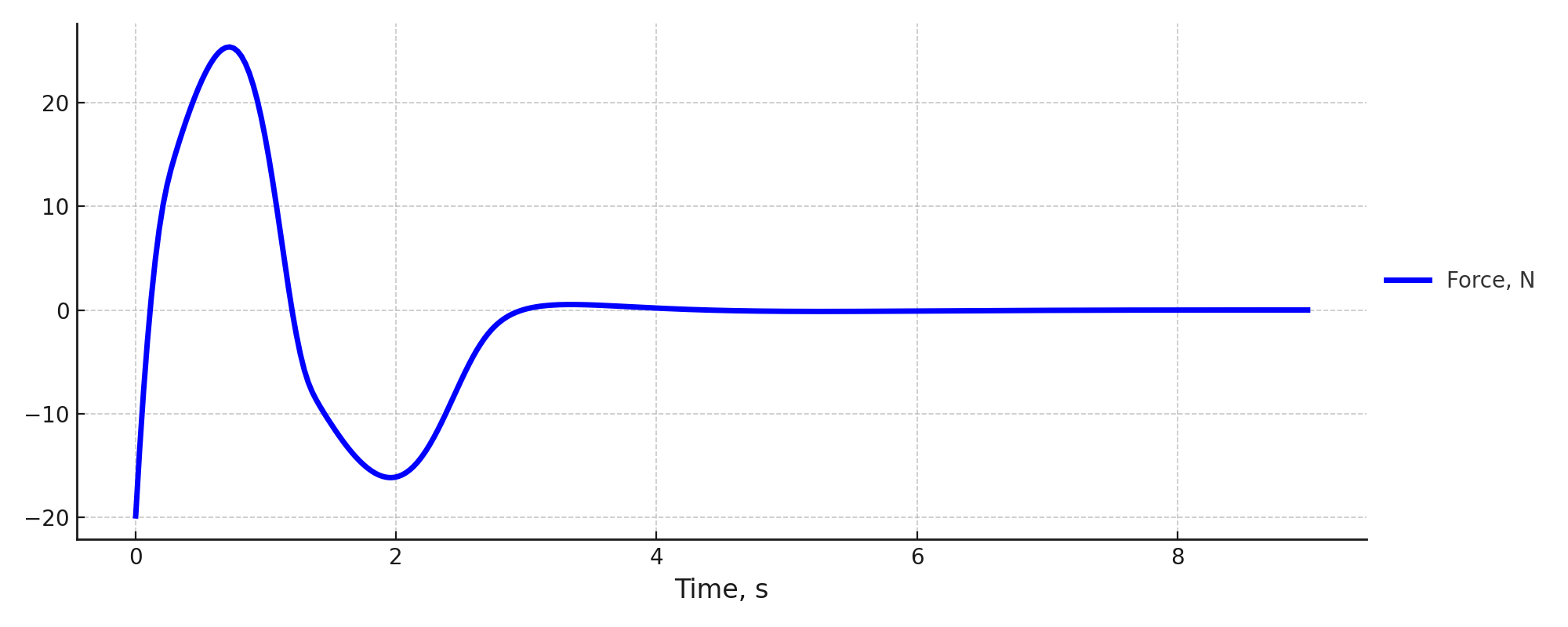}
        \caption*{Force, 1 iteration}
    \end{minipage}
    \hfill
    \begin{minipage}[b]{0.48\textwidth}
        \centering
        \includegraphics[width=\textwidth]{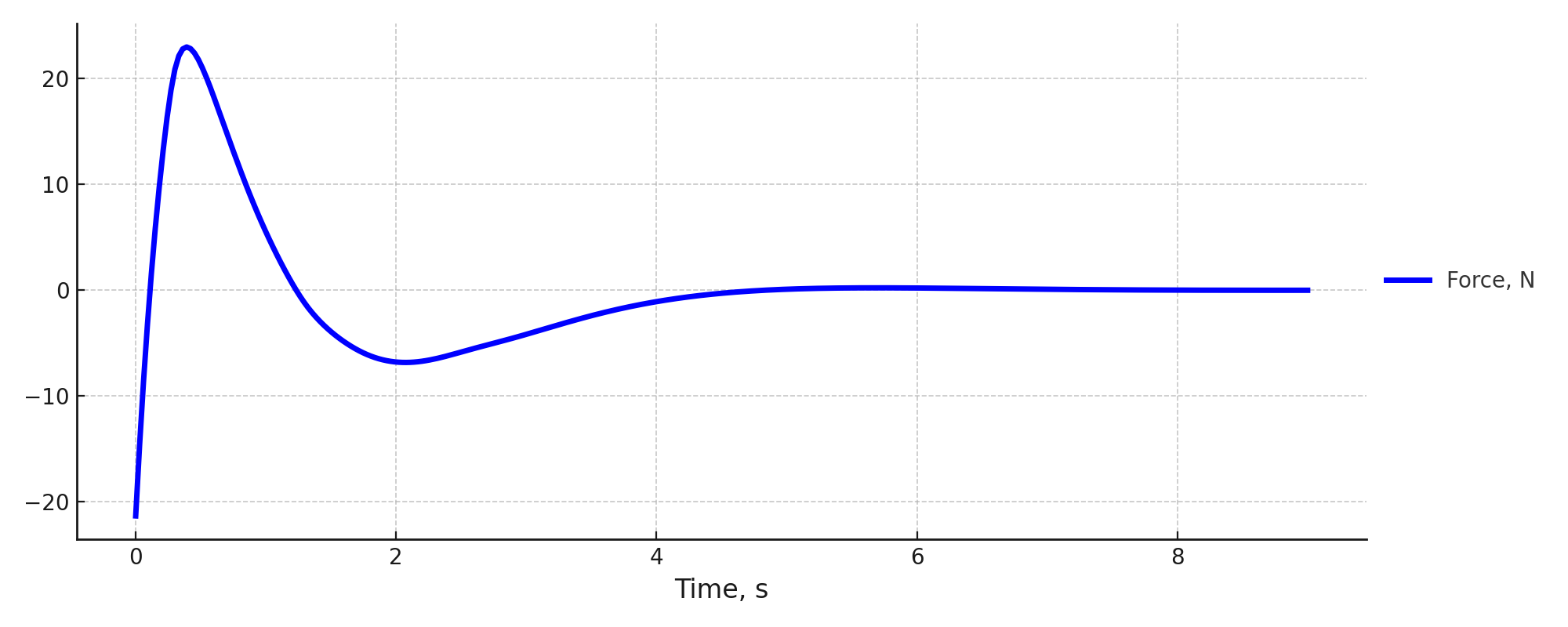}
        \caption*{Force, 3 iterations}
    \end{minipage}

    \vspace{4mm}

    \begin{minipage}[b]{0.48\textwidth}
        \centering
        \includegraphics[width=\textwidth]{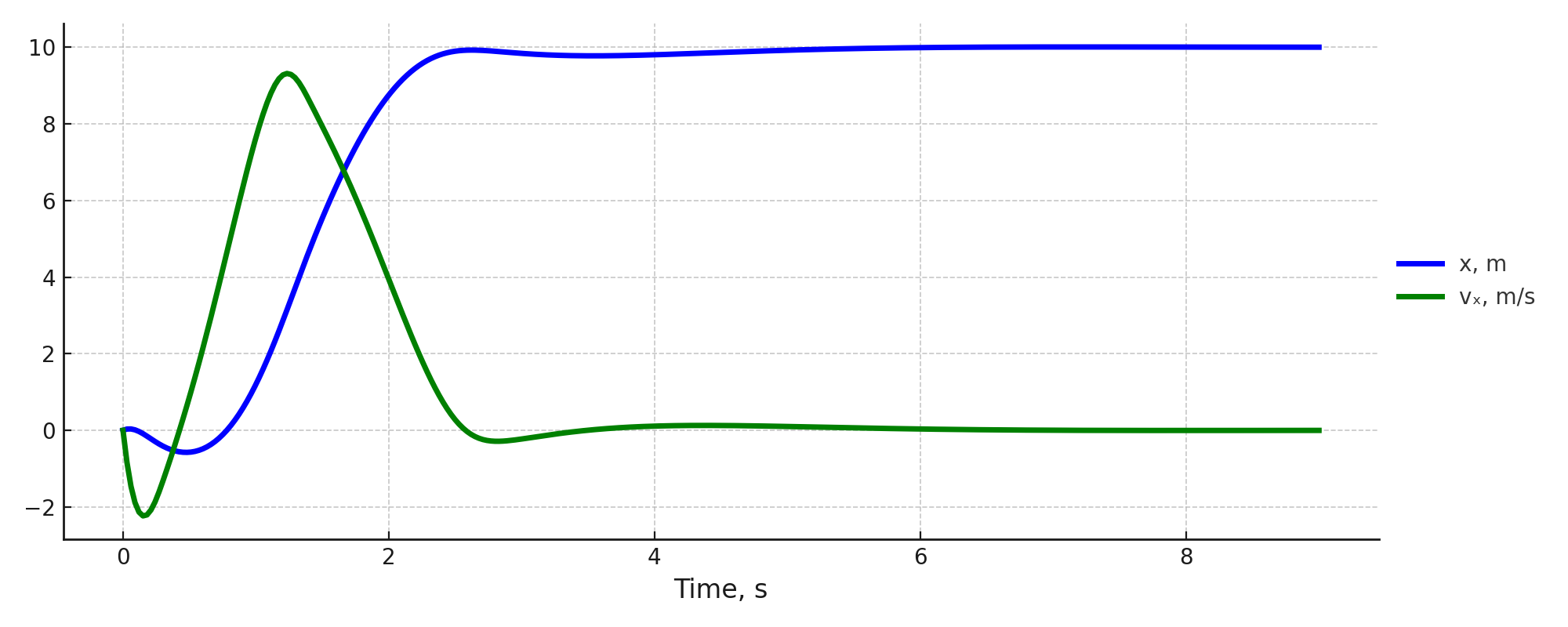}
        \caption*{Position and velocity, 1 iteration}
    \end{minipage}
    \hfill
    \begin{minipage}[b]{0.48\textwidth}
        \centering
        \includegraphics[width=\textwidth]{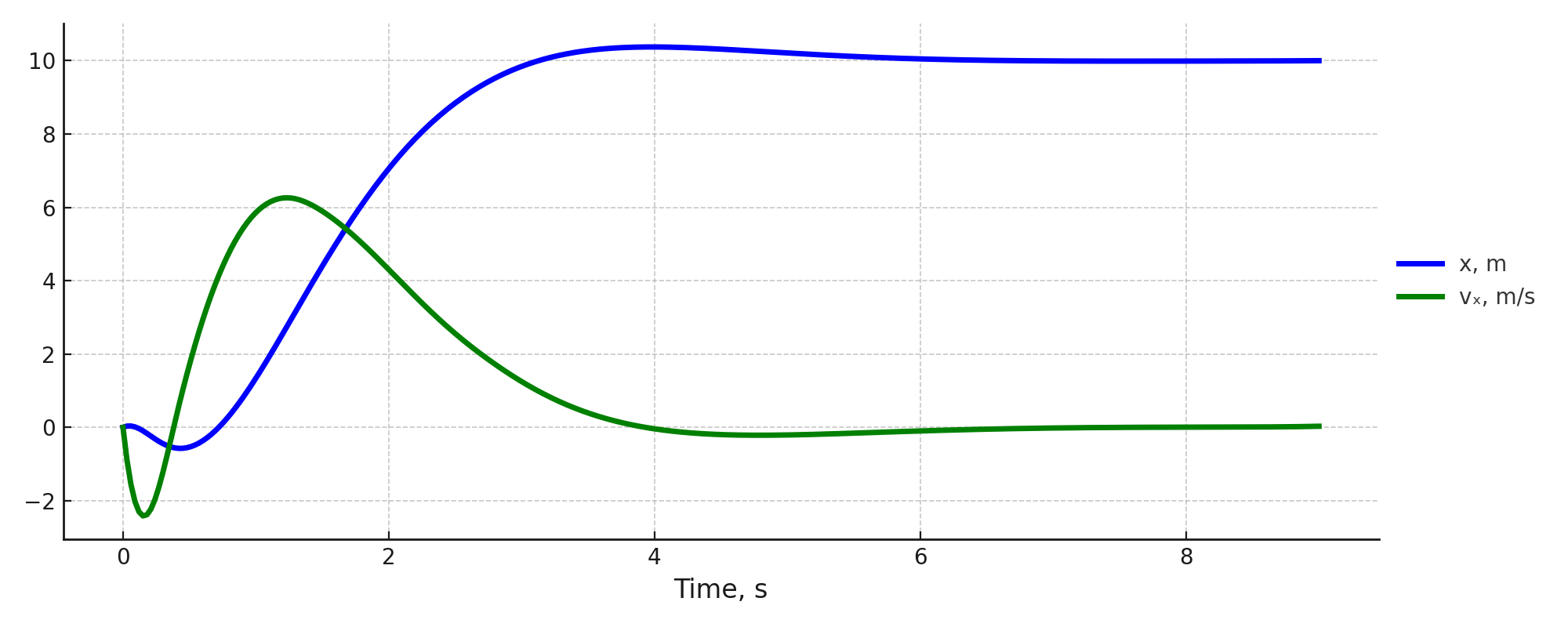}
        \caption*{Position and velocity, 3 iterations}
    \end{minipage}

    \vspace{4mm}

    \begin{minipage}[b]{0.48\textwidth}
        \centering
        \includegraphics[width=\textwidth]{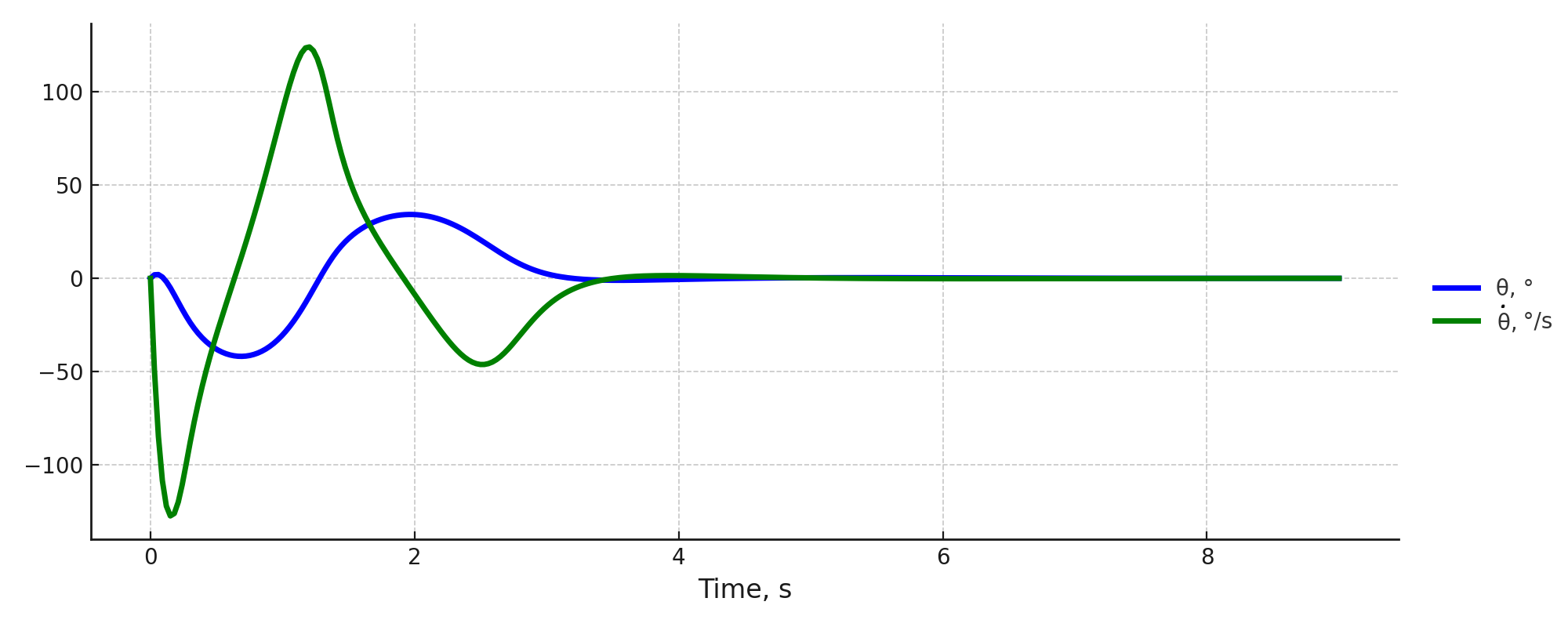}
        \caption*{Angle and angular velocity, 1 iteration}
    \end{minipage}
    \hfill
    \begin{minipage}[b]{0.48\textwidth}
        \centering
        \includegraphics[width=\textwidth]{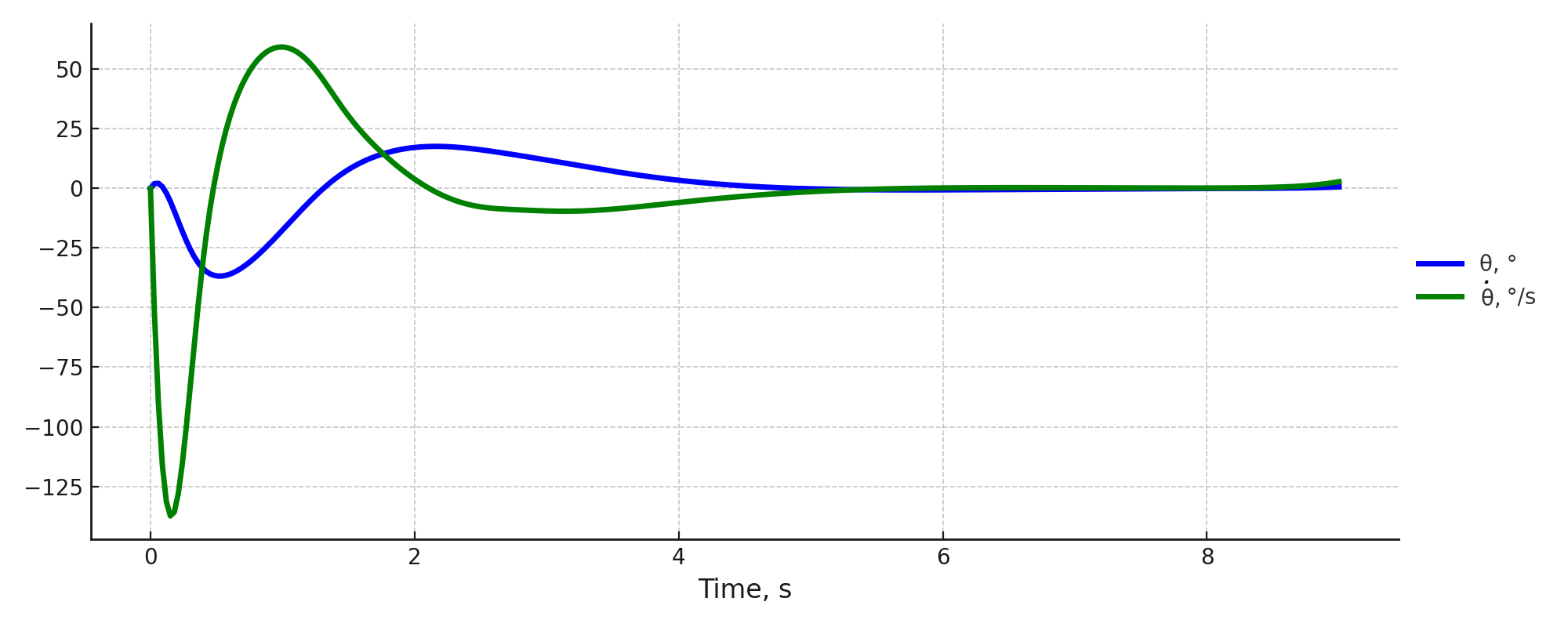}
        \caption*{Angle and angular velocity, 3 iterations}
    \end{minipage}

    \caption{Comparison of pendulum system behavior under different scenarios.}
    \label{fig:pendulum_comparison}
\end{figure}

\noindent
Figure~\ref{fig:pendulum_comparison} presents the control results using constant desired state, control inputs, and weight values for a single iteration (left column) and for three iterations (right column).

\newpage

\begin{table}[H]
\captionsetup{justification=raggedright, singlelinecheck=false}
\caption{The Single Iteration Case}
\label{tab:pendulum_const}
\begin{tabular}{ll}
\toprule
\textbf{Symbol} & \textbf{Value} \\
\midrule
$c$ (closed-loop input) & $-19.84$ \\
\midrule
$F$ (feedback matrix) &
$\begin{bmatrix}
-1.984 & -3.595 & 49.87 & 12.83
\end{bmatrix}$ \\
\bottomrule
\end{tabular}
\end{table}

\begin{figure}[H]
    \centering

    \begin{minipage}[b]{0.40\textwidth}
        \centering
        \includegraphics[width=\textwidth]{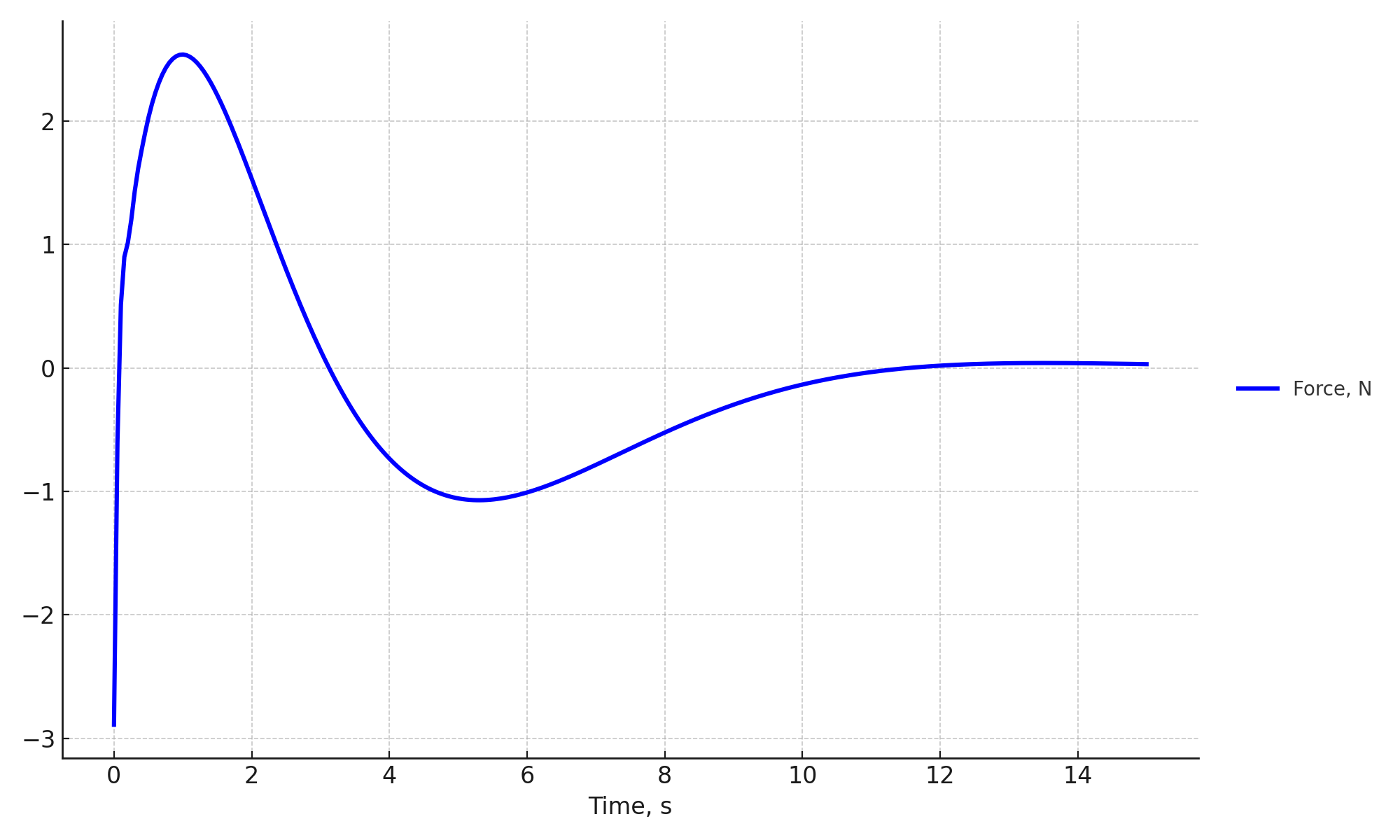}
        \caption*{Force, 1 iteration}
    \end{minipage}

    \vspace{4mm}

    \begin{minipage}[b]{0.40\textwidth}
        \centering
        \includegraphics[width=\textwidth]{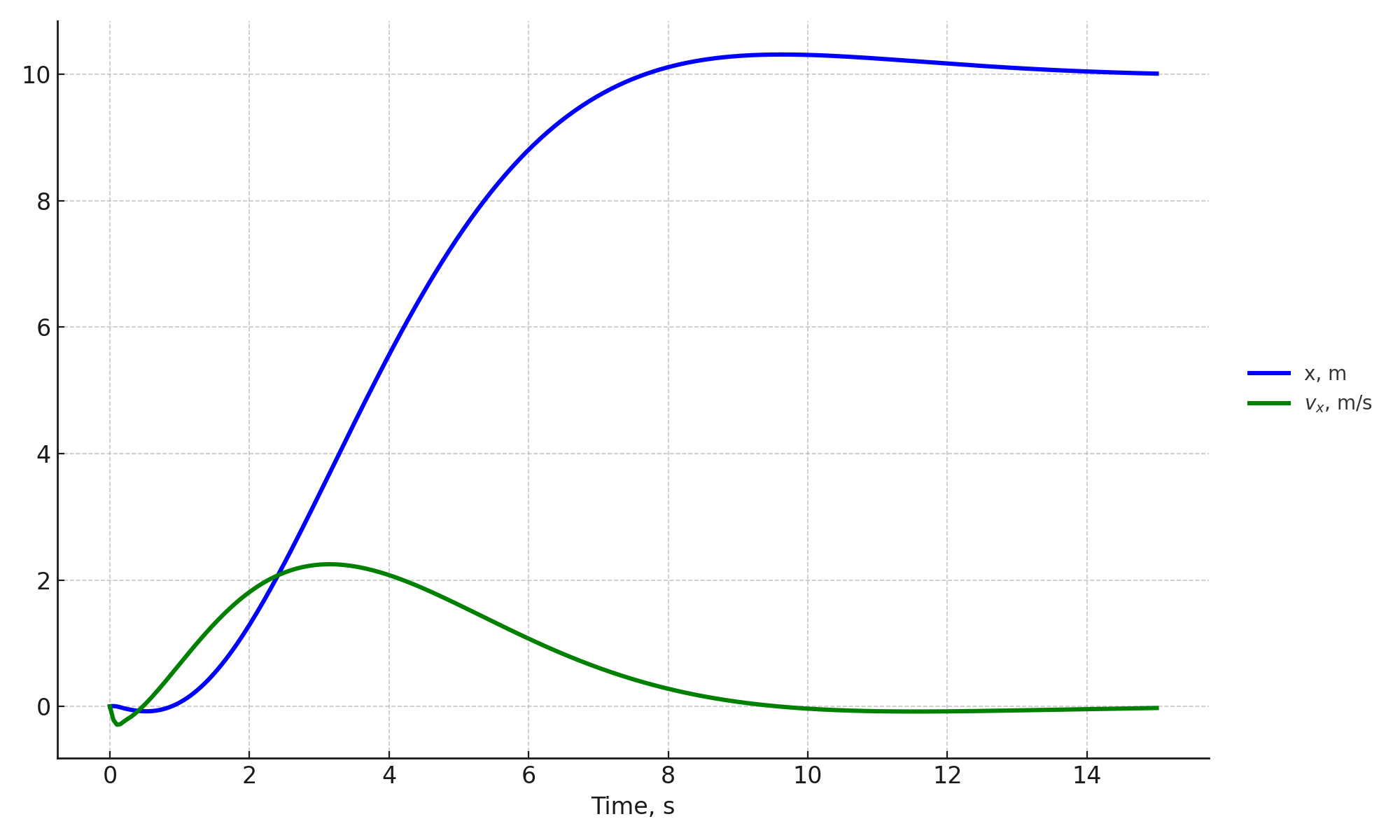}
        \caption*{Position and velocity, 1 iteration}
    \end{minipage}

    \vspace{4mm}

    \begin{minipage}[b]{0.40\textwidth}
        \centering
        \includegraphics[width=\textwidth]{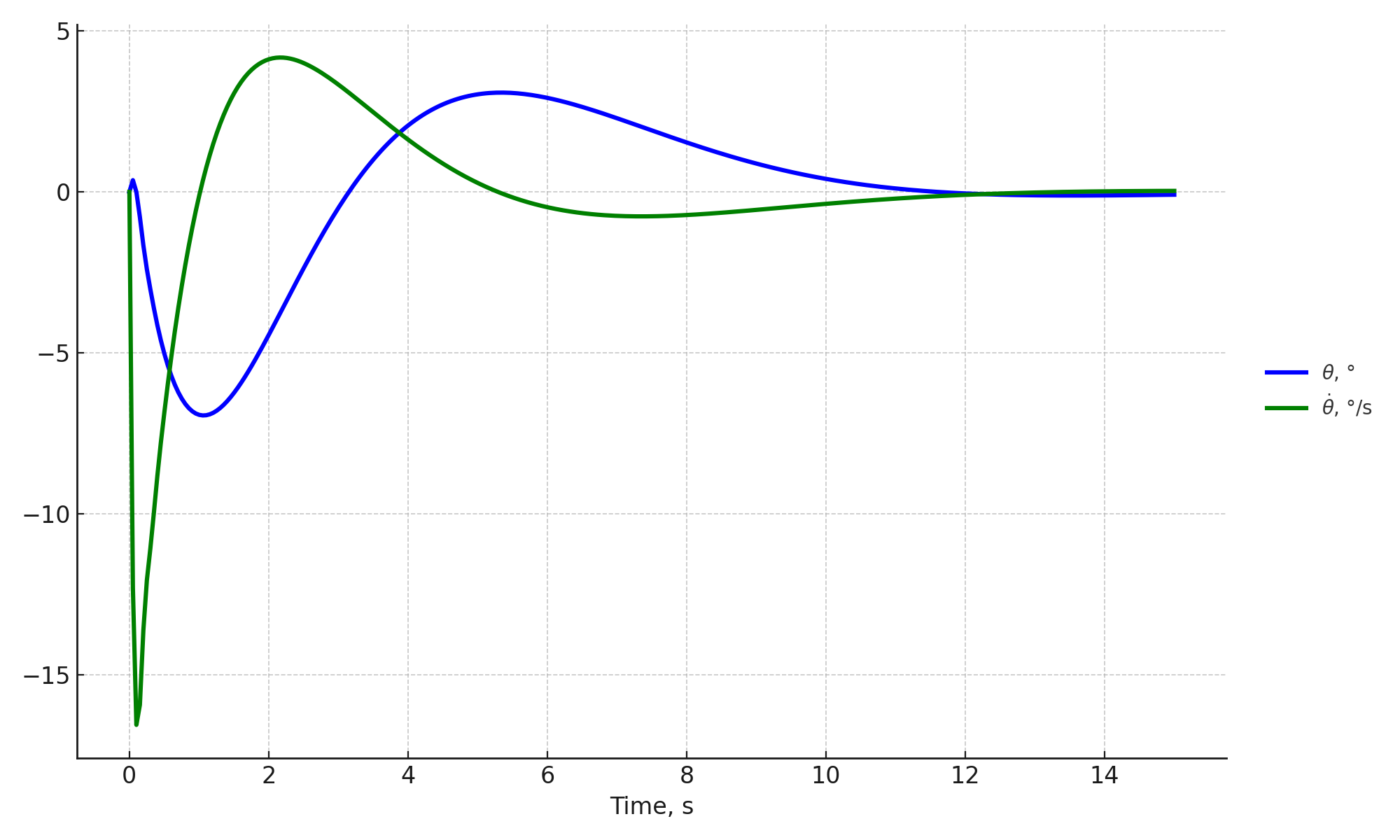}
        \caption*{Angle and angular velocity, 1 iteration}
    \end{minipage}

    \caption{Dynamic response of the pendulum system under softened control input.}
    \label{fig:pendulum_plots_2}
\end{figure}

\noindent
Another selection of weight gains, presented in Table~\ref{tab:pendulum_desired2}, results in a softened control input (force).
In this case, the transients (Figure~\ref{fig:pendulum_plots_2}) are significantly attenuated and require substantially more time to reach steady state.
A single iteration is sufficient here, as the variations in $\theta$ are relatively small and the system behaves almost linearly.

\newpage

\begin{table}[H]
\centering
\caption{Desired Values and Weights for Softened Control Input in the Inverted Pendulum System}
\label{tab:pendulum_desired2}
\begin{tabular}{c *{4}{c} @{\hspace{6pt}}|@{\hspace{6pt}} *{2}{c}}
\toprule
\textbf{Variable} & $x$ & $\dot{x}$ & $\theta$ & $\dot{\theta}$ & $u$ \\
\midrule
\textbf{Desired Value} & 10 & 0 & 0 & 0 & 0 \\
\textbf{Weight}        & 1 & 1 & 1000 & 1000 & 0 \\
\bottomrule
\end{tabular}
\end{table}

\begin{table}[H]
\centering
\caption{The Single Iteration Case for Softened Control Input}
\label{tab:pendulum_const_2}
\begin{tabular}{ll}
\toprule
\textbf{Symbol} & \textbf{Value} \\
\midrule
$c$ (closed-loop input) & $-2.889$ \\
\midrule
$F$ (feedback matrix) &
$\begin{bmatrix}
-0.2889 & -1.103 & 38.36 & 12.86
\end{bmatrix}$ \\
\bottomrule
\end{tabular}
\end{table}

\newpage

\subsection{Two-Link Manipulator}

\begin{figure}[H]
    \centering
    \begin{minipage}[b]{0.45\linewidth}
   \centering
   \includegraphics[width=\linewidth]{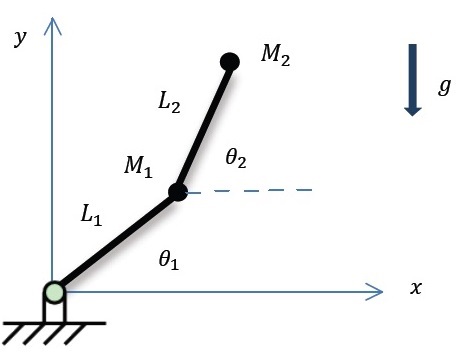}
   \end{minipage}
   \caption{Schema of Two-Link Manipulator.}
   \label{fig:TLM_schema}
\end{figure}

We consider a simplified dynamic model of a two-link manipulator depicted in Figure~\ref{fig:TLM_schema} \cite{TLM}.
The state and control variables are summarized in Table~\ref{tab:TLM_vars}, while the corresponding mathematical model is presented in Table~\ref{tab:TLM_symbol}.

\begin{table}[H]
\centering
\small
\caption{State and Control Vector for Two-Link Manipulator}
\label{tab:TLM_vars}
\begin{tabular}{llcll}
\toprule
& \# & \textbf{Variable} & \textbf{Units} & \textbf{Description} \\
\midrule
{\textbf{State Vector}} & 1 & $\theta_1$ & rad & Angle of link 1 \\
& 2 & $\dot{\theta}_1$ & rad/s & Angular velocity of link 1 \\
& 3 & $\theta_2$ & rad & Angle of link 2 \\
& 4 & $\dot{\theta}_2$ & rad/s & Angular velocity of link 2 \\
\midrule
{\textbf{Control Inputs}} & 1 & $T_1$ & Nm & Torque at joint 1 \\
& 2 & $T_2$ & Nm & Torque at joint 2 \\
\bottomrule
\end{tabular}
\end{table}

\begin{table}[H]
\centering
\small
\caption{Two-Link Manipulator Dynamic Parameters and Equations}
\label{tab:TLM_symbol}
\begin{tabular}{ll}
\toprule
\textbf{Symbol} & \textbf{Description} \\
\midrule
$g$ & Gravity constant $9.81 m/s^2$ \\
$M_1$, $M_2$ & Masses of links (each 1 kg) \\
$L_1$, $L_2$ & Lengths of links (each 1 m) \\
$m$ & $M_2/(M_1 + M_2)$ \\
$\dot{\theta}_1$ & Angular velocity of link 1 \\
$\dot{\omega}_1$ & $\frac{m T_1/(M_2L_1) - m L_2 \dot{\theta}_2^2 \sin(\theta_1 - \theta_2) - g \cos(\theta_1) - m \cos(\theta_1 - \theta_2)(T_2/(M_2L_2) + L_1 \dot{\theta}_1^2 \sin(\theta_1 - \theta_2) - g \cos(\theta_2))}{L_1(1 - m \cos^2(\theta_1 - \theta_2))}$ \\
$\dot{\theta}_2$ & Angular velocity of link 2 \\
$\dot{\omega}_2$ & $\frac{T_2/(M_2L_2) + L_1 \dot{\theta}_1^2 \sin(\theta_1 - \theta_2) - g \cos(\theta_2) - \cos(\theta_1 - \theta_2)(m T_1/(M_2L_1) - m L_2 \dot{\theta}_2^2 \sin(\theta_1 - \theta_2) - g \cos(\theta_1))}{L_2(1 - m \cos^2(\theta_1 - \theta_2))}$ \\
\bottomrule
\end{tabular}
\end{table}

\noindent
The objective is to compute an optimal control policy that drives the manipulator from zero initial conditions to the final configuration with $\theta_1 = \pi/4$ and $\theta_2 = \pi/6$ radians. The corresponding desired state and control values, along with their weights, are provided in Table~\ref{tab:TLM_desired}.

\begin{table}[H]
\centering
\small
\scriptsize
\caption{Desired Values and Weights (States vs Inputs)}
\label{tab:TLM_desired}
\setlength{\tabcolsep}{4pt}
\begin{tabular}{c *{4}{c} @{\hspace{6pt}}|@{\hspace{6pt}} *{2}{c}}
\toprule
\textbf{Variable} & $\theta_1$ & $\dot{\theta}_1$ & $\theta_2$ & $\dot{\theta}_2$ & $T_1$ & $T_2$ \\
\midrule
\textbf{Desired Value} & ${\pi}/4$ & 0 & ${\pi}/6$ & 0 & 0 & 0 \\
\textbf{Weight} & 100000 & 0 & 100000 & 0 & 1 & 1 \\
\bottomrule
\end{tabular}
\end{table}

\newpage

\begin{table}[H]
\captionsetup{justification=raggedright, singlelinecheck=false}
\caption{The Single Iteration Case}
\label{tab:TLM_const}
\begin{tabular}{ll}
\toprule
\textbf{Symbol} & \textbf{Value} \\
\midrule
$c$ (closed-loop input) &
$\begin{bmatrix}
230.3 \\
152.3\end{bmatrix}$ \\
\midrule
$F$ (feedback matrix) &
$\begin{bmatrix}
283.0 & 33.64 & 15.54 & 11.29 \\
15.53 & 11.29 & 267.4 & 22.35
\end{bmatrix}$ \\
\bottomrule
\end{tabular}
\end{table}

\begin{figure}[htbp]
    \centering

    \begin{minipage}[t]{0.48\textwidth}
        \centering
        \includegraphics[width=\linewidth]{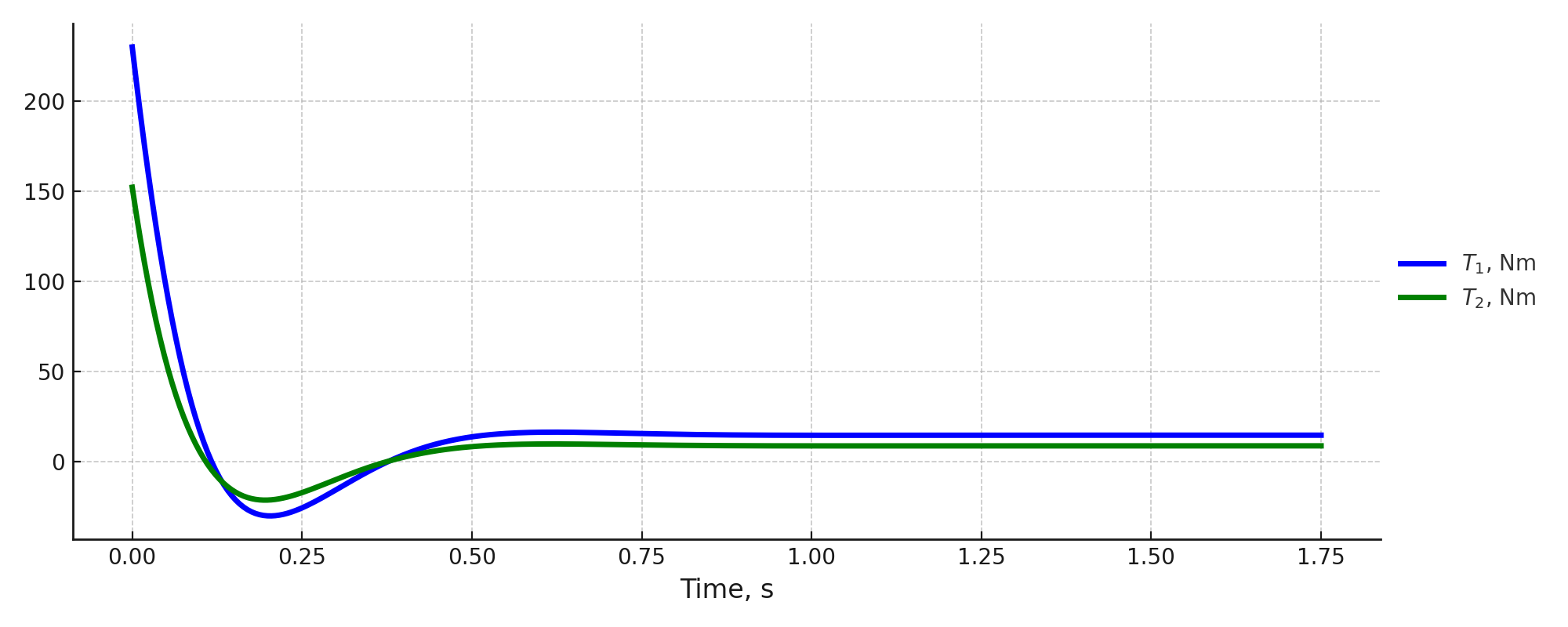}
        \caption*{Joint torques $T_1$, $T_2$, 1 iteration}
    \end{minipage}\hfill
    \begin{minipage}[t]{0.48\textwidth}
        \centering
        \includegraphics[width=\linewidth]{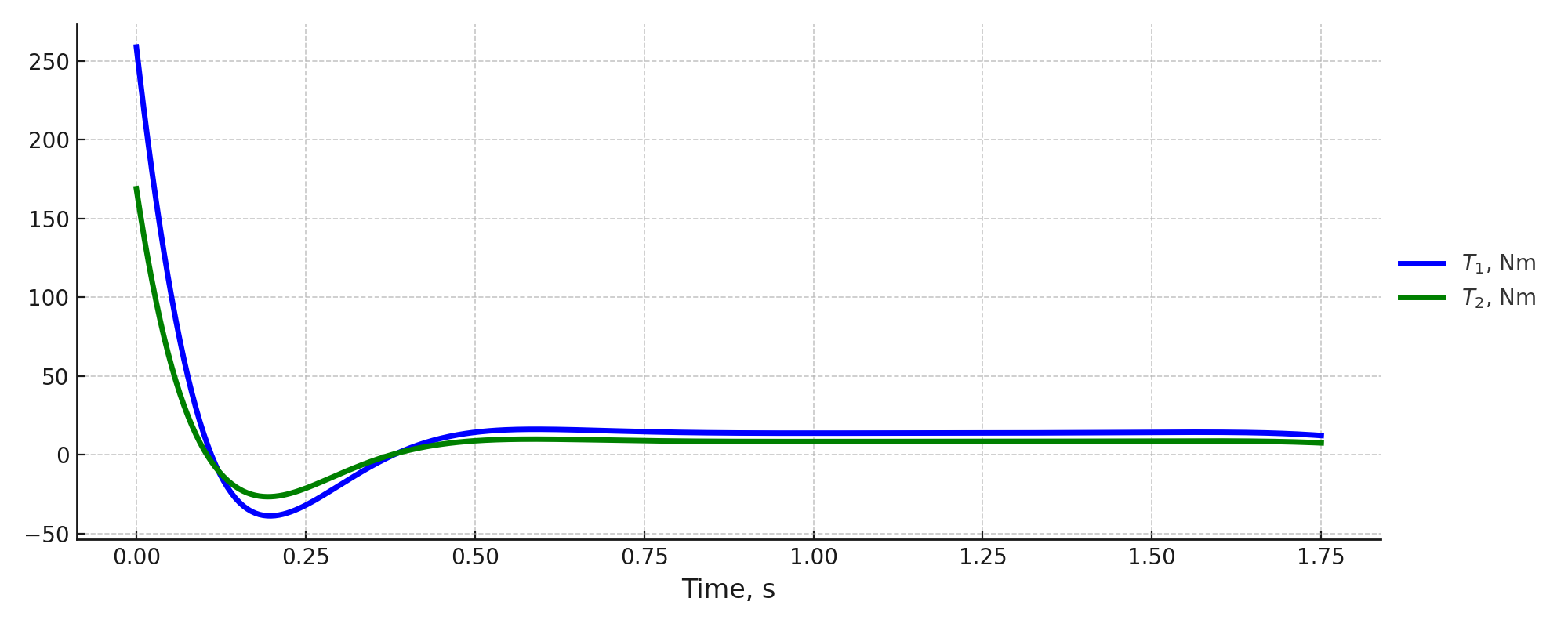}
        \caption*{Joint torques $T_1$, $T_2$, 3 iterations}
    \end{minipage}\par\medskip

    \begin{minipage}[t]{0.48\textwidth}
        \centering
        \includegraphics[width=\linewidth]{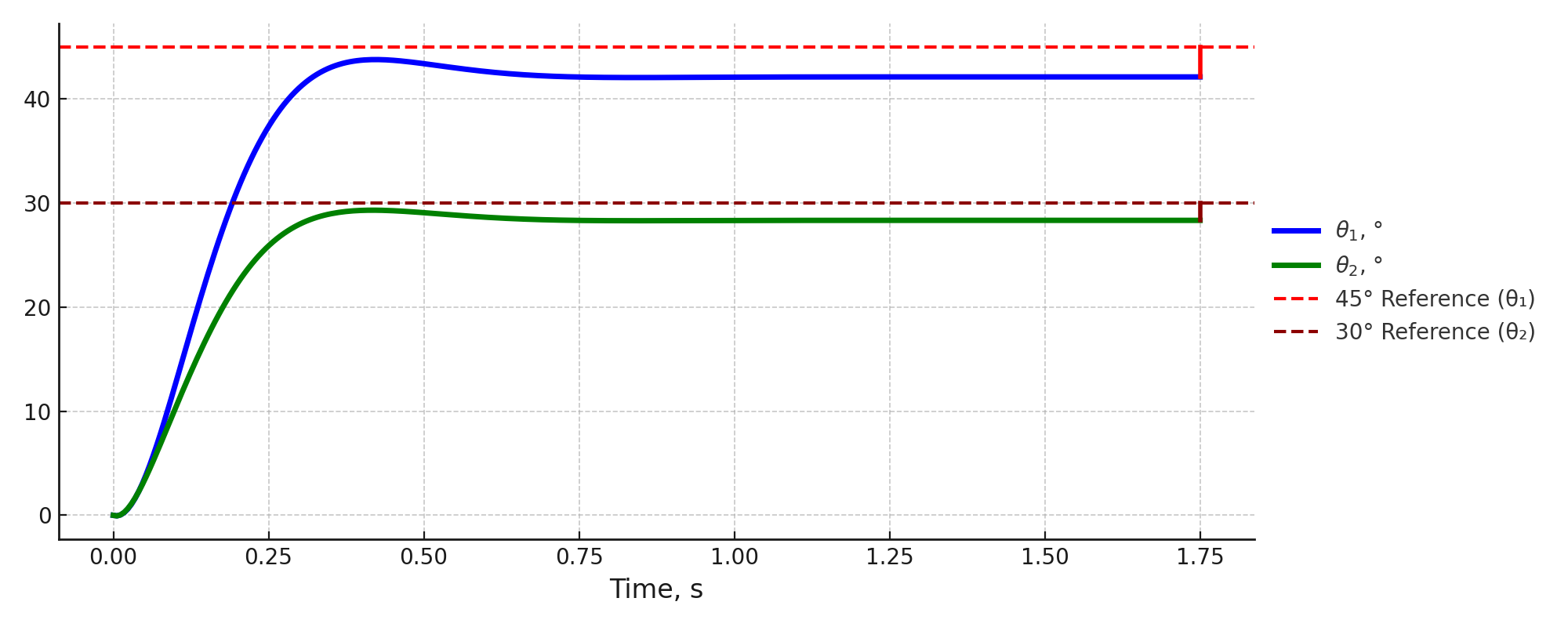}
        \caption*{Joint angles $\theta_1$, $\theta_2$, 1 iteration. Reference deviations.}
    \end{minipage}\hfill
    \begin{minipage}[t]{0.48\textwidth}
        \centering
        \includegraphics[width=\linewidth]{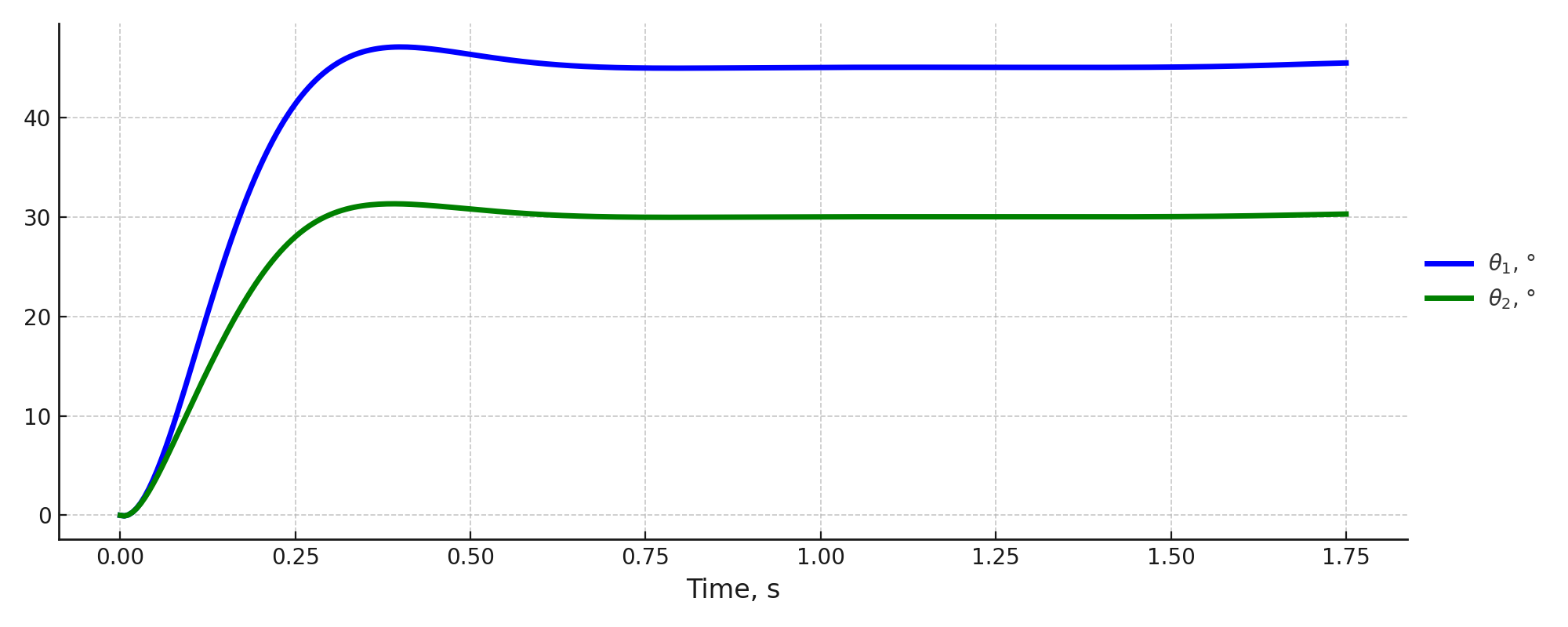}
        \caption*{Joint angles $\theta_1$, $\theta_2$, 3 iterations.}
    \end{minipage}\par\medskip

    \begin{minipage}[t]{0.48\textwidth}
        \centering
        \includegraphics[width=\linewidth]{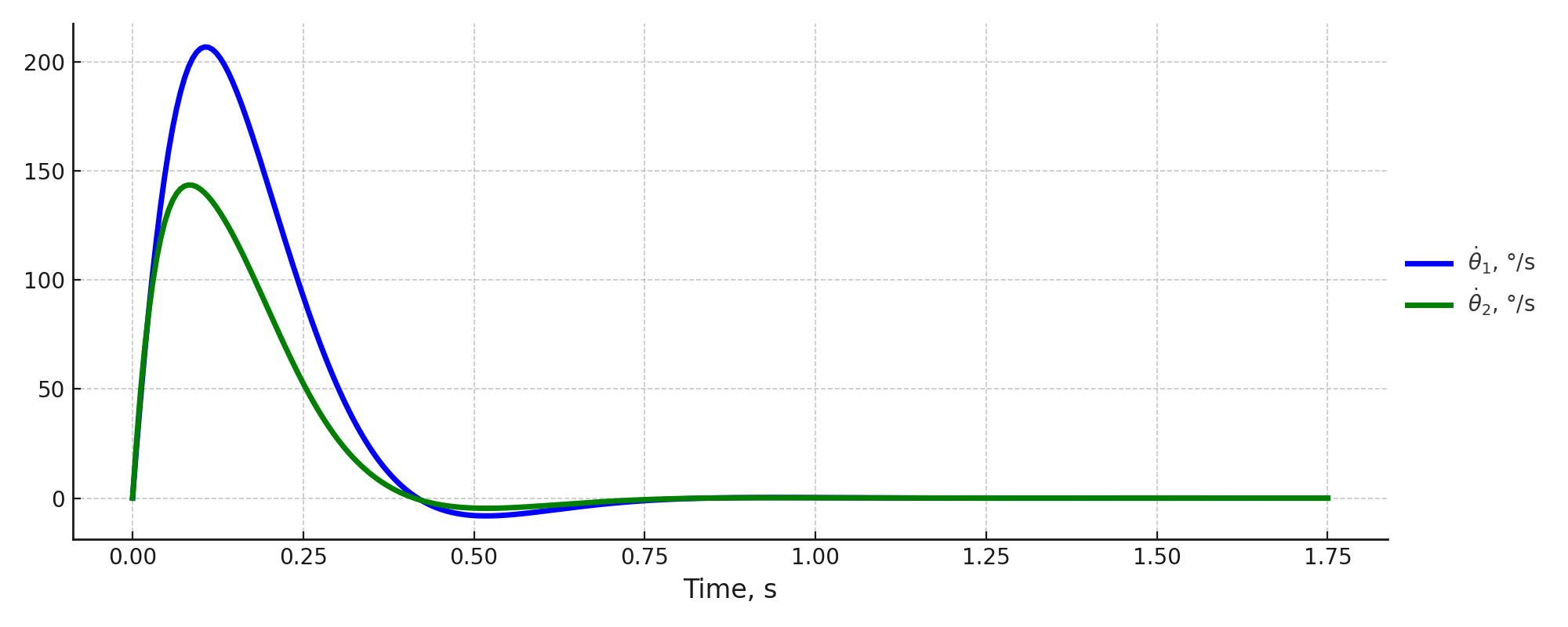}
        \caption*{Angular velocities $\dot{\theta}_1$, $\dot{\theta}_2$, 1 iteration.}
    \end{minipage}\hfill
    \begin{minipage}[t]{0.48\textwidth}
        \centering
        \includegraphics[width=\linewidth]{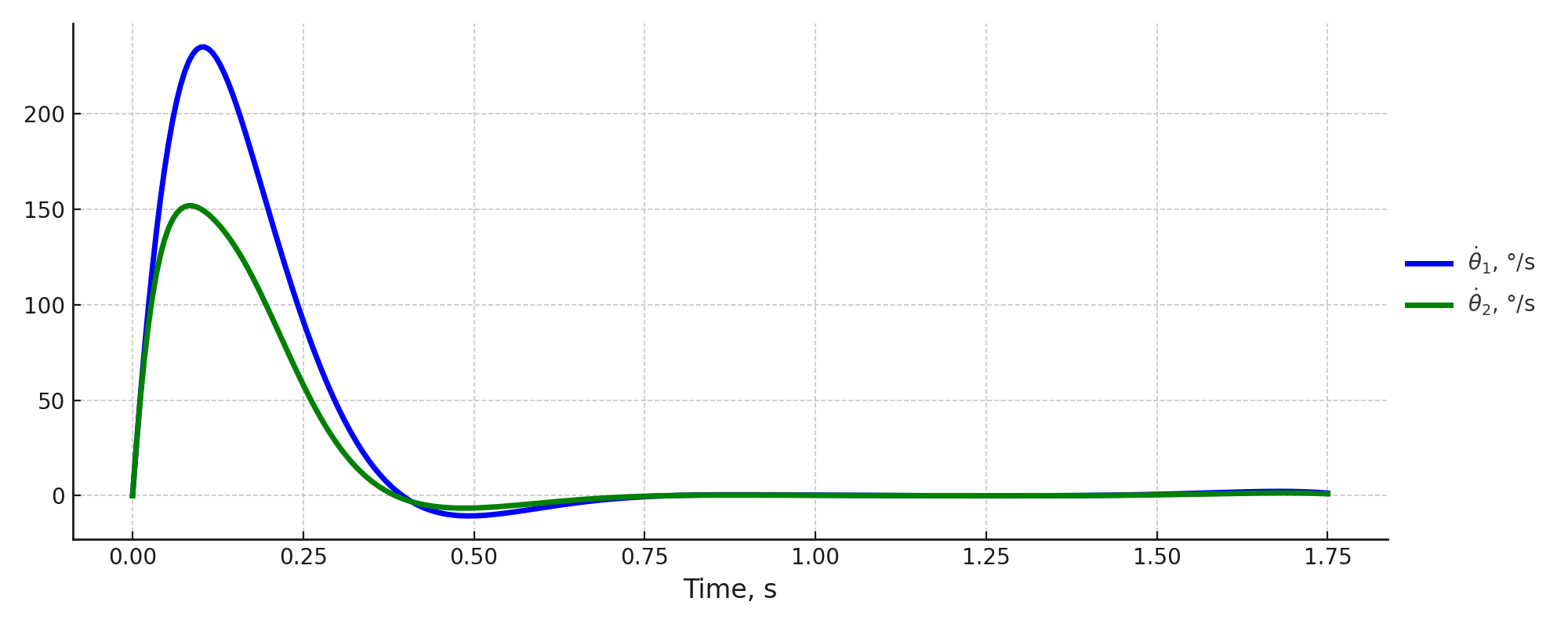}
        \caption*{Angular velocities $\dot{\theta}_1$, $\dot{\theta}_2$, 3 iterations.}
    \end{minipage}

    \caption{Two-Link Manipulator control response comparisons.}
    \label{fig:tlm-comparison}
\end{figure}

\noindent
It can be observed that under similar settings, the single-iteration case results in a steady-state deviation of the joint angles from their reference values.
Although this deviation can be corrected through parameter tuning, doing so complicates the controller design process.
In contrast, when three iterations are used, the system achieves zero steady-state error.

\newpage

\subsection{Quadcopter}

\begin{figure}[H]
    \centering
    \begin{minipage}[b]{0.45\linewidth}
    \centering
    \includegraphics[width=\linewidth]{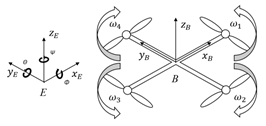}
    \end{minipage}
    \caption{Schema of Quadcopter.}
    \label{fig:Quadcopter_schema}
\end{figure}

We consider a simplified dynamic model of a quadcopter, illustrated in Figure~\ref{fig:Quadcopter_schema}~\cite{quadcopter1, quadcopter2}, which includes integrators in the control input channels to enhance control smoothness.
The associated state and control variables are listed in Table~\ref{tab:Quadcopter_vars}, and the corresponding dynamic equations are detailed in Table~\ref{tab:Quadcopter_symbol}.
The numerical values of the model parameters used in the simulations are provided in Table~\ref{tab:Quadcopter_params}.

\begin{table}[H]
\centering
\small
\caption{State and Control Vectors for the Quadcopter}
\label{tab:Quadcopter_vars}
\begin{tabular}{l c c c l}
\toprule
& \# & Variable & Units & Description \\
\midrule
\textbf{State Vector}
& 1  & $\phi$         & rad     & Roll angle \\
& 2  & $\dot{\phi}$   & rad/s   & Roll rate \\
& 3  & $\theta$       & rad     & Pitch angle \\
& 4  & $\dot{\theta}$ & rad/s   & Pitch rate \\
& 5  & $\psi$         & rad     & Yaw angle \\
& 6  & $\dot{\psi}$   & rad/s   & Yaw rate \\
& 7  & $x$            & m       & Position along $x$-axis \\
& 8  & $v_x$          & m/s     & Velocity along $x$-axis \\
& 9  & $y$            & m       & Position along $y$-axis \\
& 10 & $v_y$          & m/s     & Velocity along $y$-axis \\
& 11 & $z$            & m       & Position along $z$-axis \\
& 12 & $v_z$          & m/s     & Velocity along $z$-axis \\
& 13 & $T$            & N       & Total thrust \\
& 14 & $\tau_\phi$    & Nm      & Torque around $x$-axis (roll) \\
& 15 & $\tau_\theta$  & Nm      & Torque around $y$-axis (pitch) \\
& 16 & $\tau_\psi$    & Nm      & Torque around $z$-axis (yaw) \\
\midrule
\textbf{Control Inputs}
& 1 & $u_T$        & N/s     & Input to thrust integrator \\
& 2 & $u_\phi$     & Nm/s    & Input to $\tau_\phi$ integrator \\
& 3 & $u_\theta$   & Nm/s    & Input to $\tau_\theta$ integrator \\
& 4 & $u_\psi$     & Nm/s    & Input to $\tau_\psi$ integrator \\
\bottomrule
\end{tabular}
\end{table}

\begin{table}[H]
\centering
\small
\caption{Quadcopter Dynamic Parameters and Equations}
\label{tab:Quadcopter_symbol}
\begin{tabular}{l l c l} 
\toprule
Category & Symbol & Units & Description \\
\midrule
{Auxiliary Coefficients}
& $a_1 = \dfrac{J_y - J_z}{J_x}$ & – & Gyroscopic coefficient (roll) \\
& $a_2 = \dfrac{J_z - J_x}{J_y}$ & – & Gyroscopic coefficient (pitch) \\
& $a_3 = \dfrac{J_x - J_y}{J_z}$ & – & Gyroscopic coefficient (yaw) \\
& $b_1 = \dfrac{l}{J_x}$ & m/kg·m² & Input gain (roll) \\
& $b_2 = \dfrac{l}{J_y}$ & m/kg·m² & Input gain (pitch) \\
& $b_3 = \dfrac{l}{J_z}$ & m/kg·m² & Input gain (yaw) \\
\midrule
{Rotational Dynamics}
& $\dot{\phi} = \omega_\phi$ & rad/s & Angular velocity around $x$-axis \\
& $\dot{\omega}_\phi = a_1 \omega_\theta \omega_\psi + b_1 \tau_\phi$ & rad/s² & Angular accel. (roll) \\
& $\dot{\theta} = \omega_\theta$ & rad/s & Angular velocity around $y$-axis \\
& $\dot{\omega}_\theta = a_2 \omega_\phi \omega_\psi + b_2 \tau_\theta$ & rad/s² & Angular accel. (pitch) \\
& $\dot{\psi} = \omega_\psi$ & rad/s & Angular velocity around $z$-axis \\
& $\dot{\omega}_\psi = a_3 \omega_\phi \omega_\theta + b_3 \tau_\psi$ & rad/s² & Angular accel. (yaw) \\
\midrule
{Translational Dynamics}
& $\dot{x} = v_x$ & m/s & Velocity along $x$-axis \\
& $\dot{v}_x = \frac{T}{m} (\cos\phi \sin\theta \cos\psi + \sin\phi \sin\psi)$ & m/s² & Acceleration along $x$ \\
& $\dot{y} = v_y$ & m/s & Velocity along $y$-axis \\
& $\dot{v}_y = \frac{T}{m} (\cos\phi \sin\theta \sin\psi - \sin\phi \cos\psi)$ & m/s² & Acceleration along $y$ \\
& $\dot{z} = v_z$ & m/s & Velocity along $z$-axis \\
& $\dot{v}_z = \frac{T}{m} \cos\phi \cos\theta - g$ & m/s² & Acceleration along $z$ \\
\midrule
{Control Dynamics}
& $\dot{T} = u_T$ & N/s & Thrust control rate \\
& $\dot{\tau}_\phi = u_\phi$ & Nm/s & Torque input (roll) \\
& $\dot{\tau}_\theta = u_\theta$ & Nm/s & Torque input (pitch) \\
& $\dot{\tau}_\psi = u_\psi$ & Nm/s & Torque input (yaw) \\
\bottomrule
\end{tabular}
\end{table}

\noindent
\begin{table}[H]
\small
\centering
\caption{Model Parameters}
\label{tab:Quadcopter_params}
\begin{tabular}{>{\bfseries}c c c l}
\toprule
Parameter & Value & Units & Description \\
\midrule
$g$    & $9.81$             & $\text{m/s}^2$     & Gravitational acceleration \\
$l$    & $0.23$             & $\text{m}$         & Arm length \\
$m$    & $0.65$             & $\text{kg}$        & Mass \\
$J_x$  & $7.5 \cdot 10^{-3}$ & $\text{kg} \cdot \text{m}^2$ & Moment of inertia about $x$-axis \\
$J_y$  & $7.5 \cdot 10^{-3}$ & $\text{kg} \cdot \text{m}^2$ & Moment of inertia about $y$-axis \\
$J_z$  & $1.3 \cdot 10^{-2}$ & $\text{kg} \cdot \text{m}^2$ & Moment of inertia about $z$-axis \\
\bottomrule
\end{tabular}
\end{table}

\noindent
The objective is to compute an optimal control policy that stabilizes the quadcopter (i.e., ensures hovering), starting from the initial conditions specified in Table~\ref{tab:Quadcopter_xInit}.

\begin{table}[H]
\centering
\small
\caption{Initial State Vector $x_0$}
\label{tab:Quadcopter_xInit}
\begin{tabular}{>{\bfseries}lcccccccc}
\toprule
Parameter &
$\phi$ (rad) & $\dot{\phi}$ (rad/s) &
$\theta$ (rad) & $\dot{\theta}$ (rad/s) &
$\psi$ (rad) & $\dot{\psi}$ (rad/s) &
$x$ (m) & $v_x$ (m/s) \\
\midrule
Value &
0.3 & 1 &
-0.4 & 1 &
0.2 & 1 &
0 & 1 \\
\midrule
Parameter &
$y$ (m) & $v_y$ (m/s) &
$z$ (m) & $v_z$ (m/s) &
$T$ (N) & $\tau_\phi$ (Nm) & $\tau_\theta$ (Nm) & $\tau_\psi$ (Nm) \\
\midrule
Value &
0 & 1 &
0 & -1 &
0 & 0 & 0 & 0 \\
\bottomrule
\end{tabular}
\end{table}

\newpage

\begin{figure}[htbp]
    \centering

    \begin{minipage}[t]{0.48\textwidth}
        \centering
        \includegraphics[width=\linewidth]{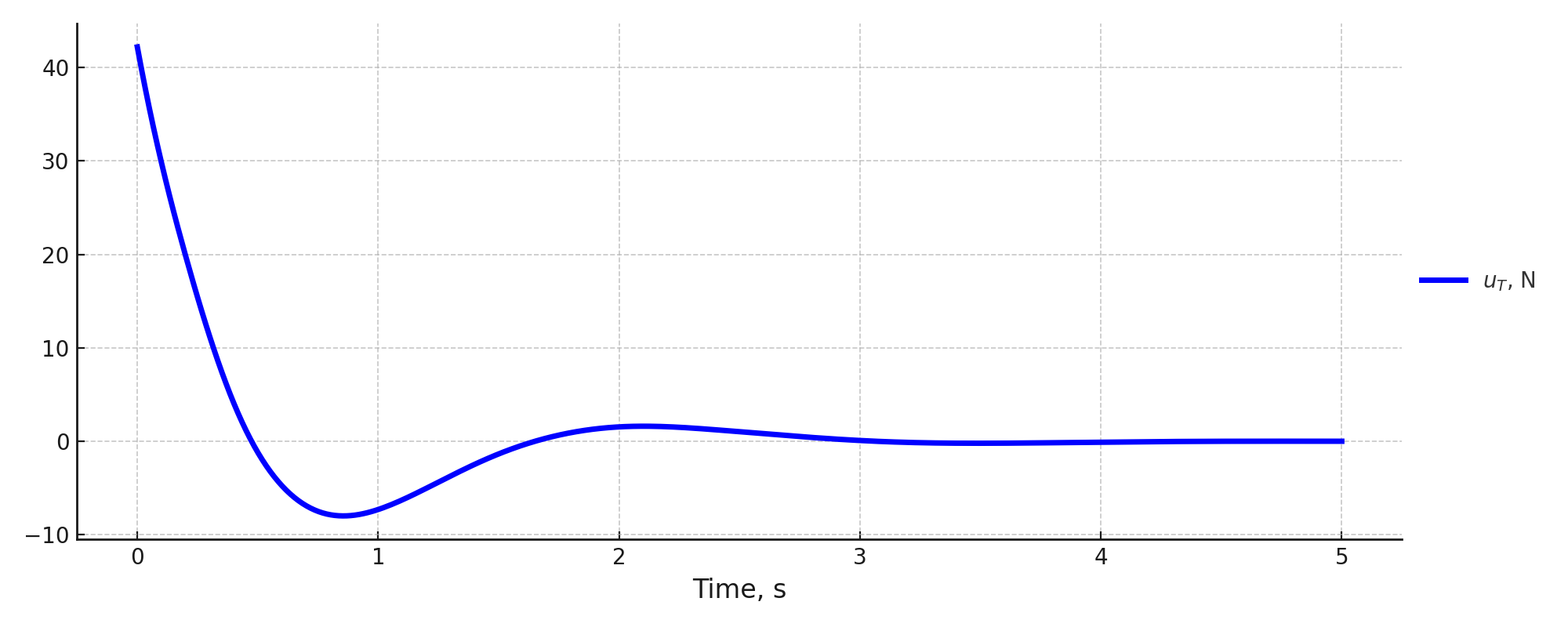}
        \caption*{Control input $u_T$}
    \end{minipage}\hfill
    \begin{minipage}[t]{0.48\textwidth}
        \centering
        \includegraphics[width=\linewidth]{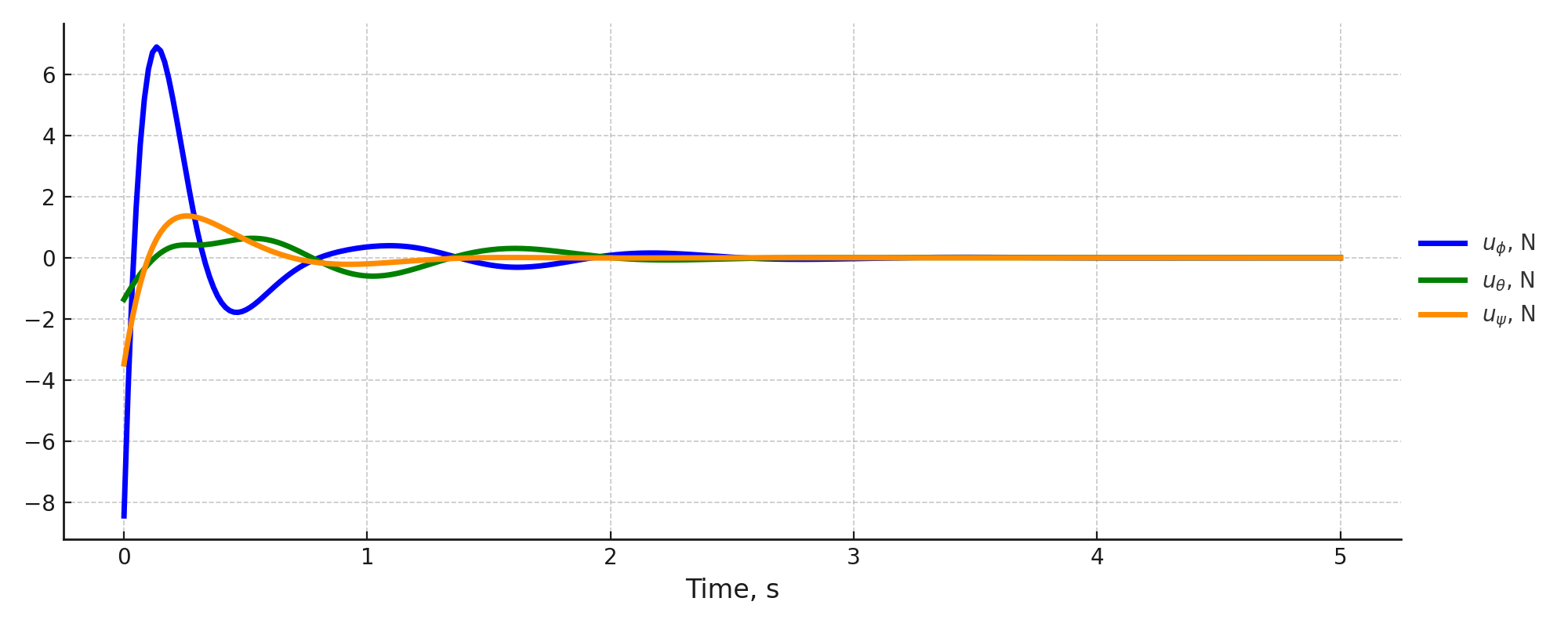}
        \caption*{Control inputs $u_\phi$, $u_\theta$, $u_\psi$}
    \end{minipage}\par\medskip

    \begin{minipage}[t]{0.48\textwidth}
        \centering
        \includegraphics[width=\linewidth]{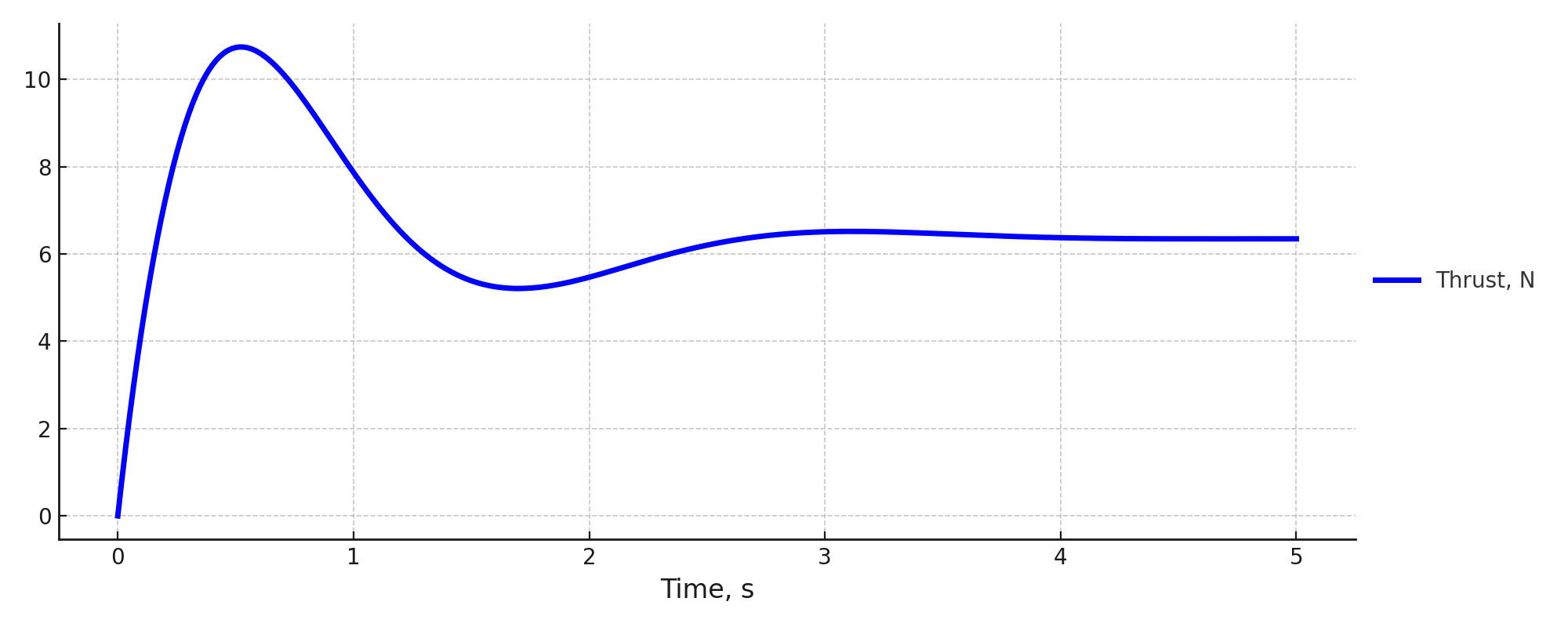}
        \caption*{Generated thrust}
    \end{minipage}\hfill
    \begin{minipage}[t]{0.48\textwidth}
        \centering
        \includegraphics[width=\linewidth]{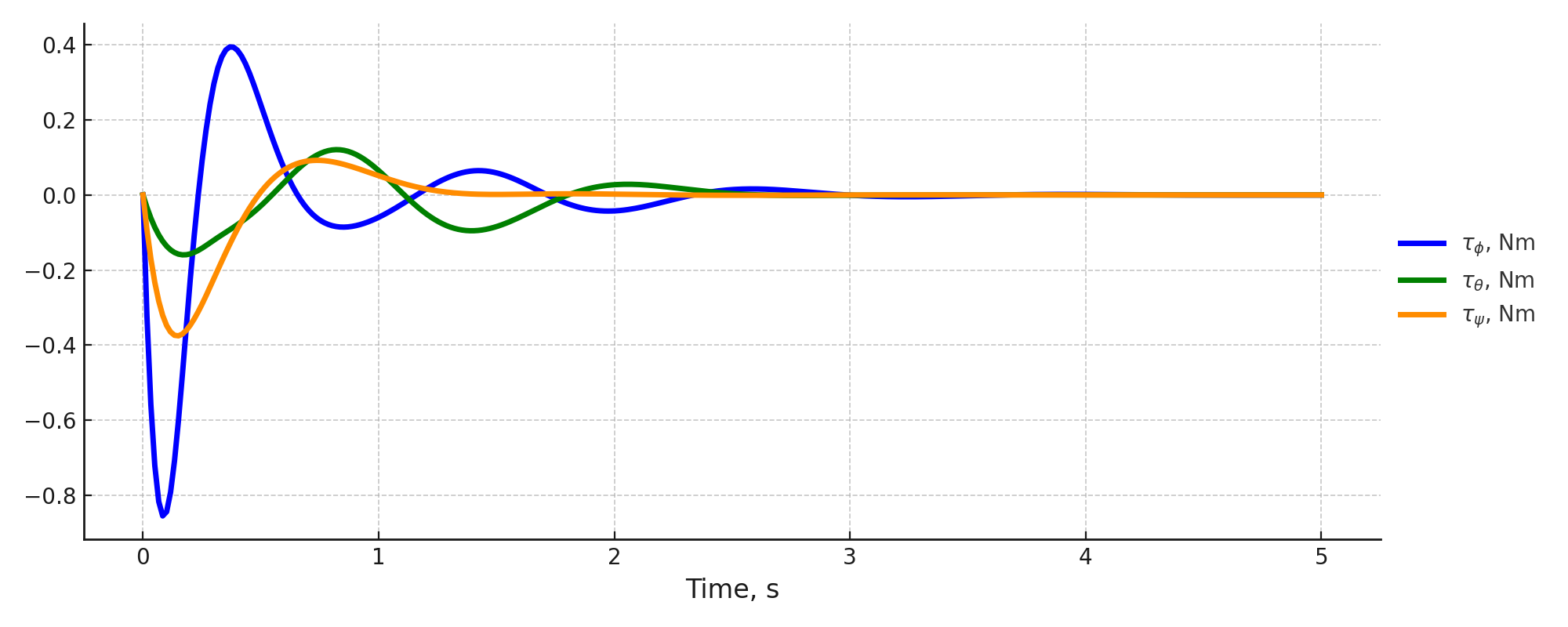}
        \caption*{Torques $\tau_\phi$, $\tau_\theta$, $\tau_\psi$}
    \end{minipage}\par\medskip

    \begin{minipage}[t]{0.48\textwidth}
        \centering
        \includegraphics[width=\linewidth]{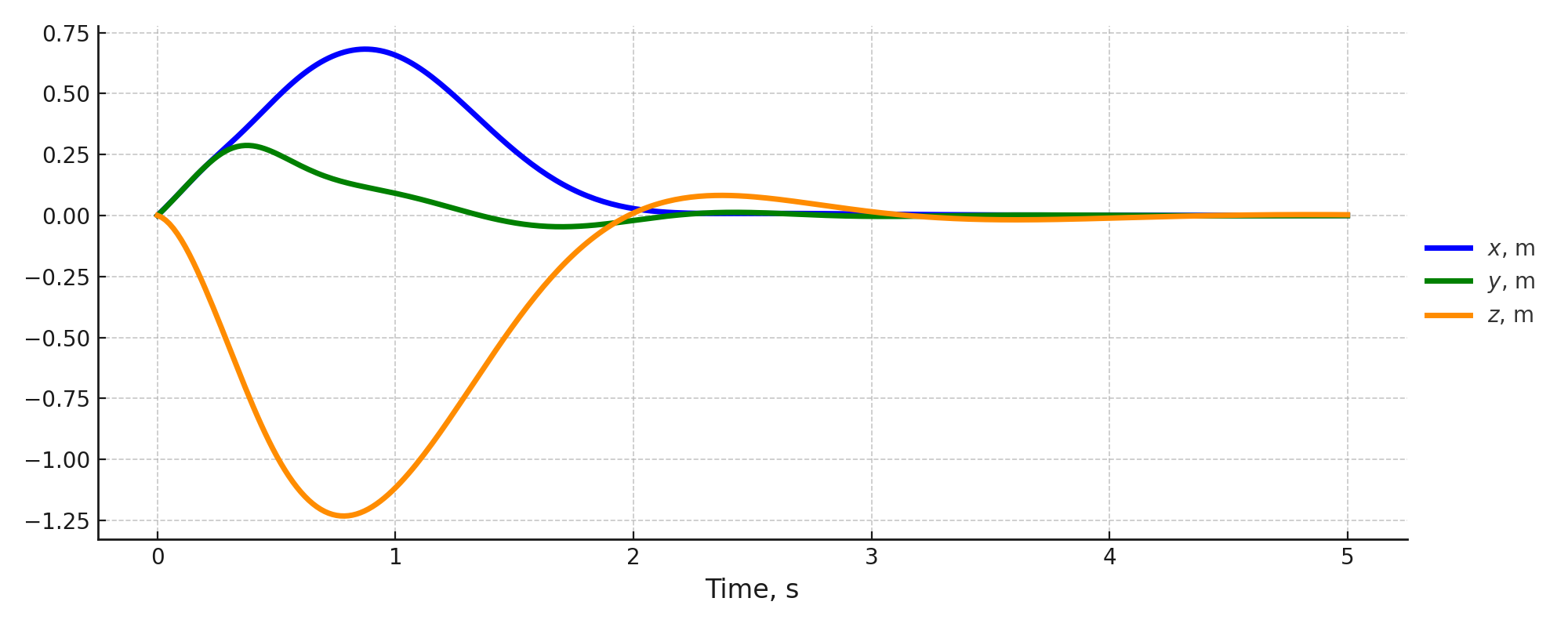}
        \caption*{Positions $x$, $y$, $z$}
    \end{minipage}\hfill
    \begin{minipage}[t]{0.48\textwidth}
        \centering
        \includegraphics[width=\linewidth]{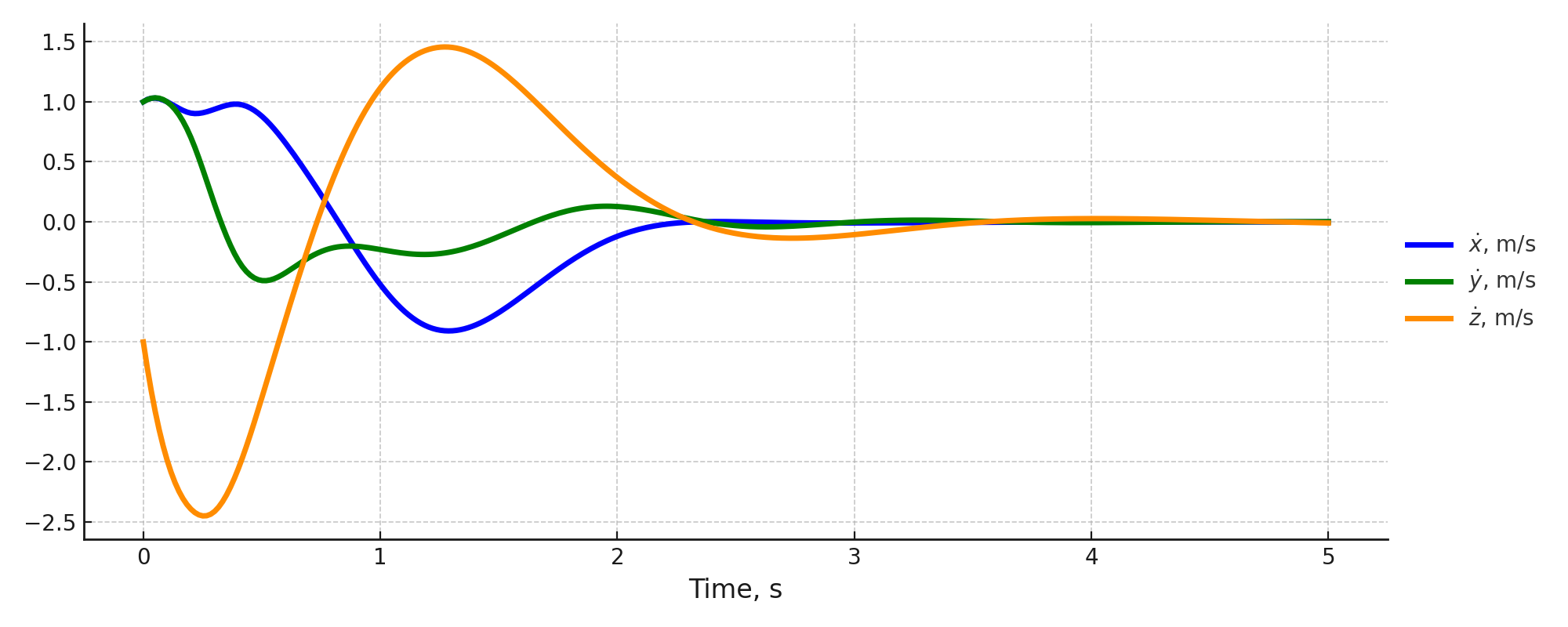}
        \caption*{Velocities $\dot{x}$, $\dot{y}$, $\dot{z}$}
    \end{minipage}\par\medskip

    \begin{minipage}[t]{0.48\textwidth}
        \centering
        \includegraphics[width=\linewidth]{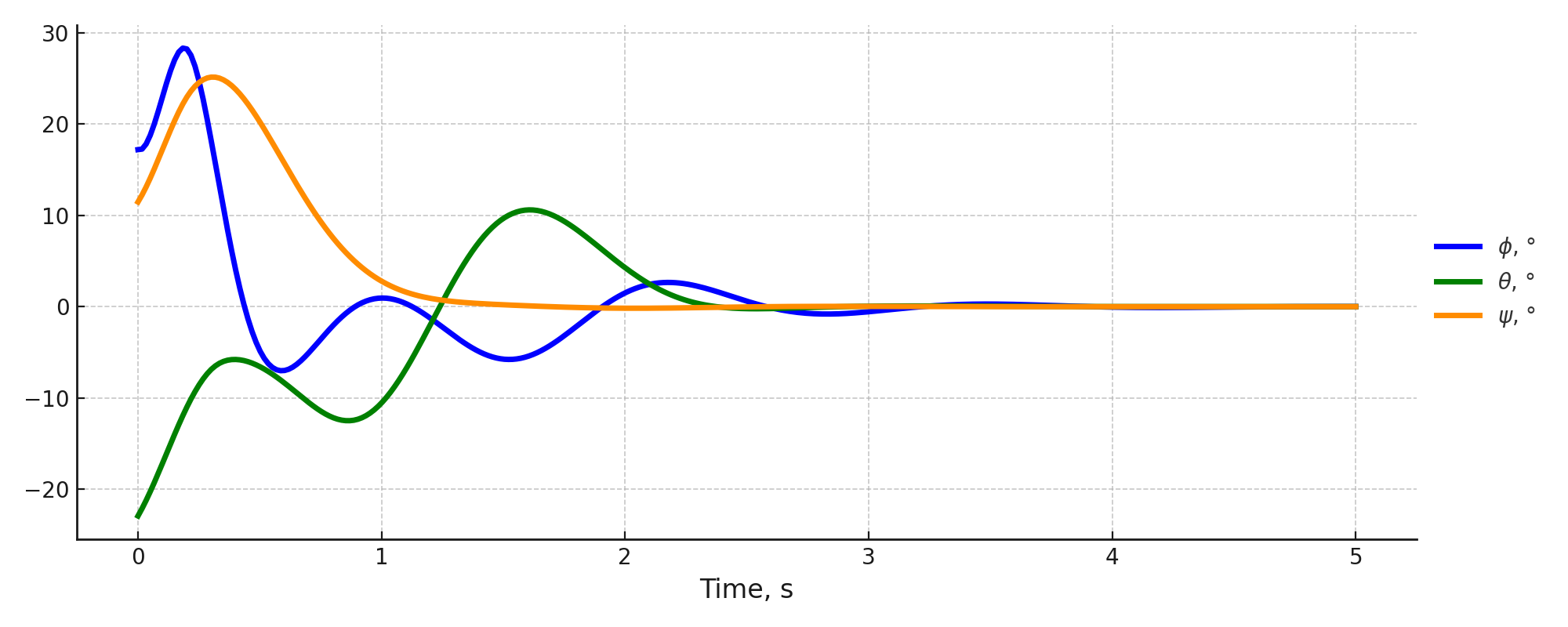}
        \caption*{Angles $\phi$, $\theta$, $\psi$}
    \end{minipage}\hfill
    \begin{minipage}[t]{0.48\textwidth}
        \centering
        \includegraphics[width=\linewidth]{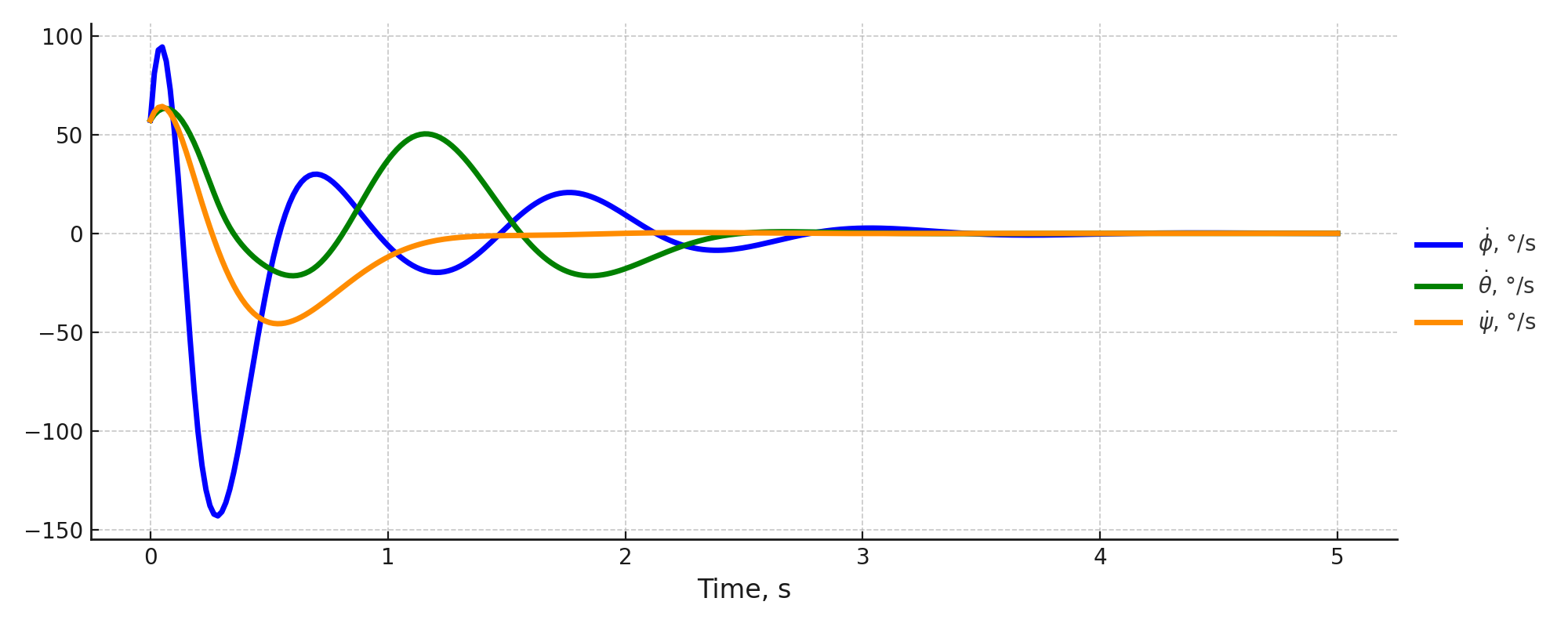}
        \caption*{Angular rates $\dot{\phi}$, $\dot{\theta}$, $\dot{\psi}$}
    \end{minipage}

    \caption{Quadcopter control and state evolution: inputs, forces, torques, position, velocity, and orientation.}
    \label{fig:quadcopter-composite}
\end{figure}

\noindent
The computational results after five iterations, using the desired state and control values along with their corresponding weights from Table~\ref{tab:Quadcopter_desired}, are presented in Figure~\ref{fig:quadcopter-composite}.

\begin{table}[H]
\centering
\scriptsize
\caption{Desired Values and Weights}
\label{tab:Quadcopter_desired}
\setlength{\tabcolsep}{4pt}
\renewcommand{\arraystretch}{1.2}
\begin{tabular}{c *{16}{c} @{\hspace{6pt}}|@{\hspace{6pt}} *{4}{c}}
\toprule
\textbf{Variable} &
$\phi$ & $\dot{\phi}$ & $\theta$ & $\dot{\theta}$ & $\psi$ & $\dot{\psi}$ &
$x$ & $v_x$ & $y$ & $v_y$ & $z$ & $v_z$ &
$T$ & $\tau_\phi$ & $\tau_\theta$ & $\tau_\psi$ &
$u_T$ & $u_\phi$ & $u_\theta$ & $u_\psi$ \\
\midrule
\textbf{Desired Value} &
0 & 0 & 0 & 0 & 0 & 0 &
0 & 0 & 0 & 0 & 0 & 0 &
0 & 0 & 0 & 0 &
0 & 0 & 0 & 0 \\
\textbf{Weight} &
10 & 1 & 10 & 1 & 10 & 1 &
10 & 1 & 10 & 1 & 50 & 5 &
0 & 0 & 0 & 0 &
0.5 & 0.5 & 0.5 & 0.5 \\
\bottomrule
\end{tabular}
\end{table}


\noindent
A single iteration, which yields constant input and feedback, is insufficient to transfer the quadcopter to the hovering position.
With two iterations, the system reaches a hovering state, though with some deviation from the initial location.
Increasing the number of iterations results in a solution that ensures final hovering precisely at the initial point.

\newpage

\section*{Conclusions}

This paper presented a unified framework for solving quadratic optimal control problems using both classical LQR and the iterative LQR (iLQR) algorithm. The approach is applicable to linear and nonlinear systems alike, leveraging discrete-time dynamic programming and system linearization.

We demonstrated the effectiveness of the iLQR approach through detailed simulations involving Rayleigh oscillator, an inverted pendulum, a two-link manipulator, and a quadcopter model.
The results confirm the method’s tracking accuracy under nonlinear conditions.

The modular structure of iLQR makes it well-suited for integration with Model Predictive Control (MPC) architectures, further enhancing its utility in real-time, constrained, or adaptive control environments. Future work may explore real-world implementation scenarios, especially for robotics and aerial vehicle control.

\end{document}